\documentclass[%
 aip,
 amsmath,amssymb,
 reprint,%
]{revtex4-1}

\usepackage{graphicx}
\usepackage{dcolumn}
\usepackage{bm}
\usepackage[utf8]{inputenc}
\usepackage[T1]{fontenc}
\usepackage{mathptmx}
\usepackage{etoolbox}
\usepackage{graphicx}
\usepackage{dcolumn}
\usepackage{bm}
\usepackage{tikz}
\usepackage{pgfplots}
\usepgfplotslibrary{colormaps}

\makeatletter
\def\@email#1#2{%
 \endgroup
 \patchcmd{\titleblock@produce}
  {\frontmatter@RRAPformat}
  {\frontmatter@RRAPformat{\produce@RRAP{*#1\href{mailto:#2}{#2}}}\frontmatter@RRAPformat}
  {}{}
}%

\begin{document}

\preprint{APS/123-QED}

\title{Activity enhances transport while competing interactions preserve structure in colloidal microphase formers}


\author{Horacio Serna\textsuperscript{$*$}}\altaffiliation{These authors equally contributed to this work}%

\affiliation{Instituto de Química Física Blas Cabrera, Consejo Superior de Investigaciones Científicas (CSIC), Calle Serrano 119, 28006 Madrid, Spain}

\author{José Martín-Roca}\altaffiliation{These authors equally contributed to this work}
\affiliation{Departament de Física de la Matèria Condensada, Universitat de Barcelona, C. Martí Franquès 1, 08028 Barcelona, Spain}

\author{Ariel G. Meyra}
\affiliation{Instituto de Física de Líquidos y Sistemas Biológicos,
CONICET-UNLP, La Plata, Argentina and Universidad Tecnológica Nacional, Facultad Regional La Plata, CIMEC - Centro de Investigación en Mecánica Experimental y Computacional, Berisso, Argentina}

\author{Eva G. Noya}
\affiliation{Instituto de Química Física Blas Cabrera, Consejo Superior de Investigaciones Científicas (CSIC), Calle Serrano 119, 28006 Madrid, Spain}

\email{hserna@iqf.csic.es}

\date{\today}

\begin{abstract}
Colloidal models with short-range attraction and long range repulsion (SALR) have been extensively studied using theoretical and simulations methods due to their rich and universal equilibrium phase behavior. Using Brownian Dynamics simulations, we study the dynamical phase behavior of active suspensions in which colloidal particles interact with each other via a SALR potential. Upon increasing the self-propulsion force of the particles, we observed that the structural transitions the active suspension undergoes resemble those observed in its passive counterpart by increasing the temperature of the thermal bath. However, when looking at the transport properties of active and passive suspensions with similar structure, we observed a clear mismatch. We demonstrated that increasing the activity enhances the particles mobility within the SALR fluid when simultaneously preserves the structure. This leads to a structure-dynamics decoupling induced by the activity whereas at the same time highlights the structural memory of SALR potentials under non-equilibrium conditions.  
\end{abstract}

\maketitle


\section{\label{sec:introduction}Introduction}

Active matter physics is a field that studies out-of-equilibrium systems, particularly focusing on ensembles of agents that are able to transform the energy, stored or supplied from the surroundings, to perform work that often manifests as self-propulsion \cite{ramaswamy2017active,gentile2023active}. Active matter systems usually exhibit exclusive collective spatio-temporal features with no equilibrium counterparts. On the one hand, collective motion emerges across a broad range of length scales and in diverse contexts, from biological microswimmers, synthetic active colloids to flocks of birds and swarms of insects \cite{vicsek2012collective}. On the other hand, Motility-Induced Phase Separation (MIPS) is arguably the most studied phenomenon in active systems at the colloidal scale, with efforts coming from theory \cite{cates2015motility,omar2023mechanical}, simulations\cite{martin2021characterization,paoluzzi2022motility,stenhammar2014phase} and experiments \cite{liu2019self,van2019interrupted,buttiononi2013dynamical}. MIPS emerges in purely repulsive active Brownian particles due to the feedback between self-trapping induced by high persistence and the mobility reduction induced by dense regions that, in turn, destabilizes the homogeneous phase leading to phase separation\cite{cates2015motility}. Although MIPS might look as an equilibrium phase separation, it violates detailed balance, thus it is an exclusive phenomenon of out-of-equilibrium systems. In practice, to observe MIPS, the system must be set at high densities and high self-propulsion speeds \cite{stenhammar2014phase}. Active colloidal suspensions at low to intermediate densities have been much less studied. Similarly, active colloids with attractive interactions have been studied comparatively much less than purely repulsive, with only a few notable studies demonstrating the formation of living clusters and active crystals \cite{prymidis2015self,mognetti2013living}.

In the equilibrium side, a colloidal model that has been extensively studied is that containing short-range attraction and long-range repulsion, the so-called SALR model \cite{imperio2006microphase,ruiz2021role,ciach2013origin,pini2017pattern,zhuang2016equilibrium}. Typically, at the colloidal scale, the short-range attraction is the result of depletion forces whereas the repulsion comes from screened electrostatic interactions \cite{ruiz2021role,ciach2013origin}. SALR interactions might also be resolved by suspending non-magnetic microparticles in a fluid containing magnetic nanoparticles, then actuated by external fields to induce the formation of clusters with a well-defined shape \cite{al2022dual}. When tuned correctly, both the pairwise potential and the thermodynamic conditions, the competing interactions of SALR models allow a rich phase behavior in the shape of density modulated phases such as cluster-crystals, hexagonal cylindrical, ordered bicontinuous and lamellar phases. Remarkably, these ordered microphases are similar to those observed in block-copolymers \cite{bates1990block} and water-oil-surfactant mixtures \cite{alexandridis1998record}. Despite the physical origin of the interactions, or the mathematical shape of the potential, it has been demonstrated that systems with competing interactions share a universal phase behavior from the structural point of view \cite{ciach2013origin}. Thus, the structural features observed in a given system with competing interactions might be useful to predict the structure of a different one under similar conditions. Since SALR models are relatively easy to treat theoretically and \textit{in silico}, they have become a cornerstone for studying self-assembly in soft materials.

Although SALR models are paradigmatic in colloidal suspensions, and the active matter community has been focused on MIPS at the colloidal scale, there are few works on active SALR systems. For example, by means of Brownian Dynamics (BD) simulations, the phase behavior of an active two-dimensional SALR suspension was studied, revealing the formation of living clusters \cite{tung2016micro}. Additionally, in a recent work, Cao et al \cite{cao2024conceptual} showed experimentally and through BD simulations how SALR interactions result in mesoscopic patterns and large-scale assembled structures with potential applications in the design of smart materials.

In this paper, using BD simulations, we study a three-dimensional system of self-propelled particles interacting via a SALR potential capable of self-assembling ordered structures such as cluster-crystal, hexagonal cylindrical and lamellar microphases. We equilibrate these phases at low temperature to achieve both short and long-range order. Then, we gradually increase the self-propulsion force to investigate the activity-induced phase transitions structurally and dynamically and compare them with their passive counterparts. We take as a reference the equilibrium phase diagram reported in \cite{serna2021formation} by some of the present authors.

We found that the active SALR system undergoes a structure-dynamics decoupling when compared with the passive case. This decoupling is evident by the fact that at equivalent global structural features, the active and passive systems exhibit despair transport properties. A more detailed analysis reveals that the origin of the decoupling relies on the active system preserving the local ordered structure typical of low temperatures even at intermediate to high activities when at the same time improves particles mobility. 

The article is organized in the following way. In Section \ref{sec:models_and_methods}, we present the model and the analysis tools we use for the simulation study. Section \ref{sec:results} contains the results and Section \ref{sec:Conlcusions} the conclusions. Finally, in Section \ref{sec:SI} we provide a summary of the Supplementary Information and a brief description of its content.

\section{\label{sec:models_and_methods}Models and methods}

\subsection{Model and simulation conditions}

The system consists of $N = 16000$ spherical particles with a diameter $\sigma$ in a cubic three-dimensional box of side length $L$. We simulate the system in bulk, so periodic boundary conditions (PBC) are applied along all three axes. We consider Active Brownian Particles (ABP) following overdamped translational and rotational dynamics, 

\begin{equation}
    \frac{d \textbf{r}_i}{dt} = \frac{D_T}{k_BT}\left(- \sum_{j\neq i}\boldsymbol{\nabla}_i u_{_\text{SALR}}(r_{ij}) +  F_a\textbf{n}_i \right) + \sqrt{2D_T}\boldsymbol{\Lambda}^T_i
    \label{e:BrownianTranslation}
\end{equation}

\begin{equation}
    \frac{d \textbf{n}_i}{dt} =\sqrt{2D_R} \boldsymbol{\Lambda}^{R}_{i}\times\textbf{n}_i
    \label{e:BrownianRotational}
\end{equation}

Here $\textbf{r}_i$ is the position vector and $\textbf{n}_i$ the unit orientation vector of the particles. $F_a$ is the magnitude of the active force that acts along the orientation of the particles. $D_T$ is the translational diffusion coefficient in the dilute regime, $k_B$ is the Boltzmann constant, and $T$ is the temperature of the thermal bath. $D_R$ is the rotational diffusion coefficient that follows the relationship $D_R =3D_T/\sigma^2$, according to the Stokes-Einstein relation for spherical particles with rotational diffusion. The thermal noise is modeled via the vectors, $\boldsymbol{\Lambda}_i^T$ and $\boldsymbol{\Lambda}_i^R$, with zero-mean Gaussian white noise components with variance, $\langle \Lambda_{i,\alpha}^T (t) \Lambda_{j,\beta}^T (t') \rangle = \delta_{ij} \delta_{\alpha\beta} \delta(t-t')$,  $\langle \Lambda_{i,\alpha}^R (t) \Lambda_{j,\beta}^R (t') \rangle = \delta_{ij} \delta_{\alpha\beta} \delta(t-t')$, where $\delta_{ij}$ is the Kronecker delta, and $\delta(t)$ the Dirac distribution. Indexes $\alpha$ and $\beta$ run in the Cartesian components ${x,y,z}$, and $i$ and $j$ in the particles of the system.

The interaction between particles is modelled with an effective SALR (Short-range Attraction Long-range Repulsion) potential that results from the addition of a Lennard-Jones (LJ) 12–6 term and a screened electrostatic interaction represented by a Yukawa potential.

\begin{equation}
 \frac{u_{_\text{SALR}}(r_{ij})}{\epsilon} = 4 \frac{\epsilon_{_{LJ}}}{\epsilon} \left[ \left(\frac{\sigma_{_{LJ}}}{r_{ij}}\right)^{12}  - \left(\frac{\sigma_{_{LJ}}}{r_{ij}}\right)^{6}\right] + \frac{A/\epsilon}{r_{ij}/\lambda}e^{\left(-r_{ij}/\lambda\right)}.
 \label{e:USALR}
 \end{equation}

\begin{figure}[h!]
    \centering
    \includegraphics[width = 0.45\textwidth]{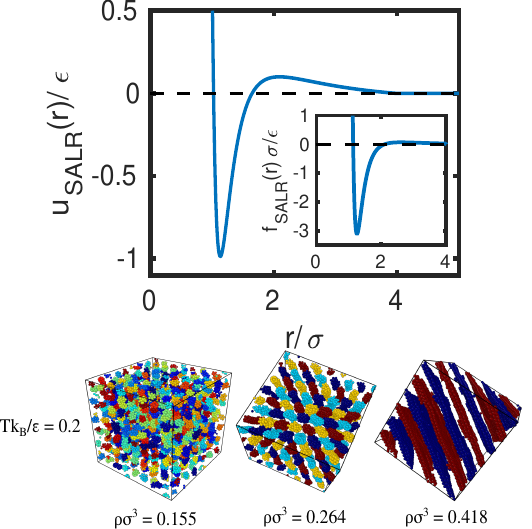}
    \caption{\textit{Interaction potential and structure of the ordered microphases.} Top panel: The Short-range Attraction and Long-range Repulsion (SALR) interaction pair potential between particles. The inset presents the associated force $f_{SALR}(r)\sigma/\epsilon = -\nabla u_{SALR}(r)/\epsilon$. At equilibrium, the system self-assembles, in order of increasing number density ($\rho\sigma^3$), into a cluster-crystal (bottom-left), cylindrical hexagonal (bottom-central) and lamellar phases (bottom-right) at low enough temperatures. Particles belonging to the same cluster are depicted with the same color for clarity.}
    \label{fig:SALR}
\end{figure}

 Here $r_{ij}$ is the distance between particles $i$ and $j$, $\epsilon_{LJ}$ and $\sigma_{_{LJ}}$ are the usual LJ parameters, $A$ determines the strength of the screened electrostatic interaction, and $\lambda$ is the Debye screening length. We choose $\epsilon_{LJ}=1.6 \epsilon$, $\sigma_{_{LJ}}=1.0 \sigma$ , $A = 0.65 \epsilon$, $\lambda = 2.0\sigma$, values for which cluster-crystal, hexagonal cylindrical, and lamellar phases form as reported in \cite*{serna2021formation}. The potential is truncated and shifted at $r_c = 4.0\sigma$. The top panel of Figure \ref{fig:SALR} presents a plot for the SALR potential and its inset contains the associated force. All the quantities are expressed in reduced dimensionless units with $\epsilon$, $\sigma$ and $m$ (mass of a particle) as fundamental units, being $\tau = \sqrt{m\sigma^2/\epsilon}$, the time unit.

Before introducing activity into the system, we first equilibrate it to reach a steady equilibrium state. To do this, we start from a random configuration of $N=2000$ particles and follow the same protocol as in Ref. \cite{chacon2022intrinsic}. First, we perform Molecular Dynamics (MD) simulations in the NpT ensemble for $1.5\cdot 10^7$ steps with a time step of $\Delta t_1 = 10^{-3}\tau$, allowing the system to reach the desired temperature and pressure. After this stage, the system is allowed to relax for $5\cdot 10^6 $ time steps in the NVT ensemble. To properly adjust the thermostat fluctuations, the system then evolves following Brownian Dynamics (BD), still without activity, for $5\cdot 10^7$ steps with $\Delta t_2 = 10^{-4}\tau$. After that, we replicate the system in the three positive directions of the axes, then a new cubic box is formed whose side is twice the original, thus reducing the calculation time to achieve a fully equilibrated system with $N = 16000$ particles.

We run for $2.2\times10^{7}$ time steps of passive BD to relax the larger system. This protocol is repeated for three different number densities ($\rho\sigma^3$) to equilibrate the cluster-crystal, the hexagonal cylindrical, and the lamellar phases at $Tk_B/\epsilon = 0.20$ as shown in the bottom panel of Figure \ref{fig:SALR}. At this temperature, the three phases also exhibit intra-cluster order as discussed in \cite{serna2021formation}. We chose a low temperature to minimize thermal fluctuations and focus on the effects of activity on the system's phase behavior. With such a value of the temperature, the translational diffusion coefficient in the dilute regime is $D_T = 0.20 \sigma^2/\tau$ and the rotational diffusion coefficient $D_R = 0.60/ \tau$, as dictated by the fluctuation-dissipation theorem with friction coefficient,  $\gamma = 1.0m/\tau$. Finally, starting from these completely equilibrated phases, we gradually increase the self-propulsion force $F_a$ and run for $6.6\times10^7$ time steps, with $\Delta t = 10^{-4}\tau$, for each value of $F_a$. The simulation with the next $F_a$ value uses the final configuration from the previous one as the initial configuration, thereby mimicking a gradual heating process. We save configurations every $3\times10^4$ time steps from the last $3\times10^7$ time steps for each $F_a$ value to compute all the considered observables at the steady state. All simulations presented in this work were performed using the Large-scale Atomic/Molecular Massively Parallel Simulator (LAMMPS)\cite{LAMMPS}.

\subsection{Analysis tools}

The structural correlations of the system are quantified by the radial pair correlation function, $g(r)$, as follows:

\begin{equation}
g(r)=\frac{1}{\rho\sigma^3 N}\left\langle \sum_{i\neq j} \delta\!\left(r-\lvert \mathbf{r}_i-\mathbf{r}_j\rvert\right)\right\rangle ,
\label{gr}
\end{equation}

where $\rho\sigma^3$ is the number density.

To further characterize local structural ordering, we employ the Common Neighbor Analysis (CNA) method , which classifies the bonding environment around each particle based on shared-neighbor connectivity \cite{faken1994systematic}. In particular, CNA allows us to identify icosahedral local order, providing a measure of its prevalence within the system. We use the implementation provided by the visualization tool OVITO \cite{stukowski2010visualization} with a cut-off radius $r_{cna} =1.5\sigma$.

To quantify local orientational order, we compute the averaged bond–order parameter \(\bar{q}_6\) introduced by Lechner and Dellago \cite{lechner2008accurate}. For each particle \(i\), the complex coefficients

\begin{equation}
q_{6m}(i)=\frac{1}{N_b(i)}\sum_{j=1}^{N_b(i)} Y_{6m}\!\left(\hat{\mathbf{r}}_{ij}\right)\label{e:q6}
\end{equation}
are obtained by averaging the spherical harmonics $Y_{6m}$ over its $N_b(i)$ nearest neighbors , defined as those particles closer than a $1.50\sigma$ cut-off distance. The locally averaged parameter $\bar{q}_6(i)$ is then computed by including the neighbors of $i$ in the average. We set the threshold ${q}_6^{s} = 0.407$, value at which the $\bar{q}_6$ probability distributions of liquid-like and solid-like systems do not overlap, to distinguish between particles with hexagonal solid-like environments ($\bar{q}_6(i) > {q}_6^{s} $) and disordered liquid-like environments ($\bar{q}_6(i) < {q}_6^{s} $). We use this order parameter to quantify the local order of the lamellar phase at low temperature which mostly exhibit hexagonal symmetry as described in \cite{serna2021formation}.

To study the diffusion in the system, we compute the mean-squared displacement (MSD), defined as

\begin{equation}
\mathrm{MSD}(\tau)/\sigma^2=\frac{1}{N}\sum_{i=1}^{N}
\left\langle \left| \mathbf{r}_i(\tau+\tau_0)-\mathbf{r}_i(\tau_0) \right|^{2} \right\rangle_{\tau_0},
\label{e:msd}
\end{equation}

where the average is taken over all particles and over different time lags $\tau_0$. In the long-time diffusive regime, the MSD grows linearly with time, and the effective diffusion coefficient $D_{\small\text{eff}}\tau/\sigma^2$ is obtained from the Einstein-Smoluchowski relation

\begin{equation}
D_{\small\text{eff}}= \frac{1}{6}\, \lim_{t \to \infty} \; \frac{\mathrm{d}}{\mathrm{d}t}\,\mathrm{MSD}(t),
\end{equation}
providing a direct measure of transport in the system.

The mechanical response of the active system is characterized by computing the effective pressure proposed by Winkler, Wysocki and Gompper \cite{winkler2015virial}. In this framework, the effective pressure, $p_{\small\text{eff}} \, \sigma^3/\epsilon$, contains both the standard virial contribution and an active term arising from self-propulsion.

\begin{equation}
p_{\small\text{eff}} = \rho k_{\mathrm{B}}T 
+ \frac{1}{3V}\left\langle \sum_{i<j} \mathbf{r}_{ij}\cdot\mathbf{F}_{ij} \right\rangle
+ \frac{F_a}{3V}\sum_{i = 1}^N \left\langle \sum_{i<j} \mathbf{n}_{i}\cdot\mathbf{r}_{i} \right\rangle ,
\label{e:p_eff}
\end{equation}

$\mathbf{r}_{ij}$ and $\mathbf{F}_{ij}$ are the pair separation vector and interaction force.

To characterize the thermodynamic state of the system from its microscopic configuration, we compute the configurational temperature, which relates temperature to the geometry of the potential energy landscape. Following the definition based on the hypervirial theorem
\cite{powles2005temperatures}, the configurational temperature is given by
\begin{equation}
k_{\mathrm{B}}T_{\mathrm{conf}}
= \frac{\left\langle \lvert \nabla U \rvert^{2} \right\rangle}
       {\left\langle \nabla^{2} U \right\rangle},
       \label{e:T_conf}
\end{equation}
where \(U\) is the total potential energy, \(\nabla U\) its gradient with respect to particle coordinates, and \(\nabla^{2}U\) the Laplacian of the potential. This definition provides a purely structural estimator of temperature, independent of particle velocities. $T_{\mathrm{conf}}$ has already been used to study non-equilibrium systems and it is usually acknowledged as a good definition in active systems \cite{saw2023configurationalA,saw2023configurationalB,hecht2024define}. We will elaborate more on the choice of $T_{\mathrm{conf}}$ in Section \ref{sec:results}.

\section{\label{sec:results}Results}
\subsection{Structure and dynamics}

\begin{figure*}[hbt!]
    \centering    \includegraphics[width=0.95\textwidth]{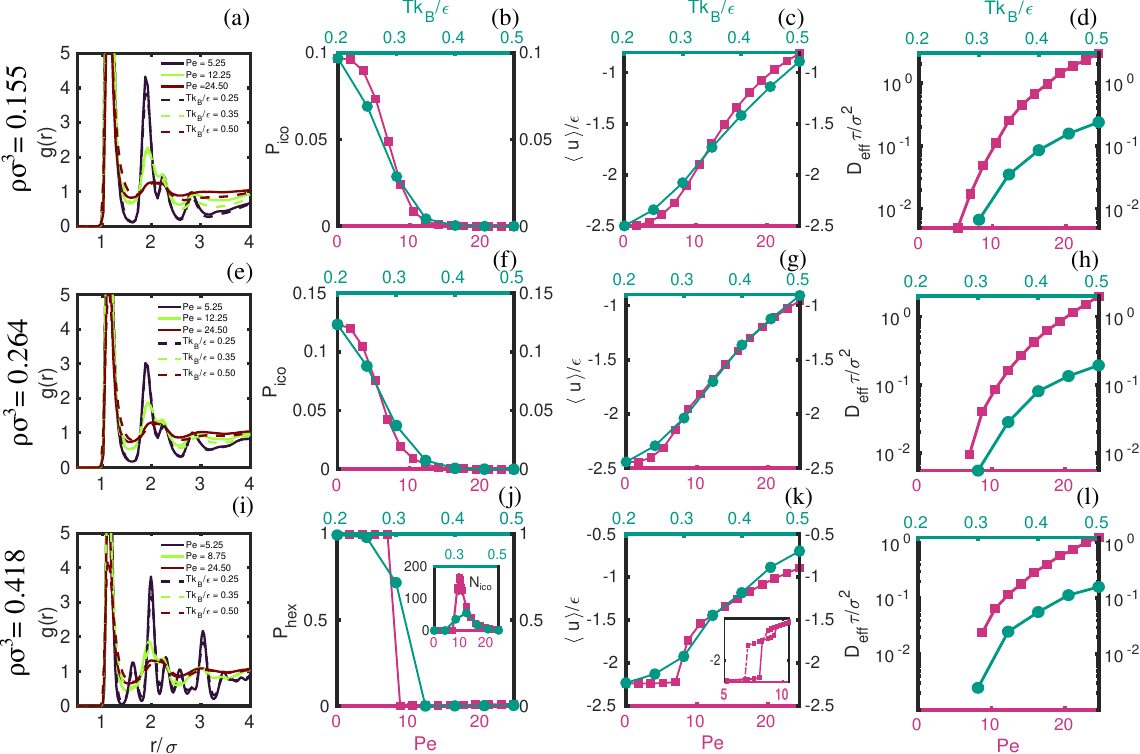}
    \caption{\textit{Comparison of structural and dynamical properties between passive and active systems.} Each row corresponds to a certain number density: $\rho\sigma^3 = 0.155$  \textbf{(a)-(d)} (cluster-crystal at low temperature/activity), $\rho\sigma^3 = 0.264$ \textbf{(e)-(h)} (hexagonal cylindrical  at low temperature/activity) and $\rho\sigma^3 = 0.418$ \textbf{(i)-(l)} (lamellar at low temperature/activity). The first column (\textbf{(a)}, \textbf{(e)} and \textbf{(i)}) contains the pair correlation functions, $g(r)$, of selected active and passive systems with similar structure for the three densities considered. Solid lines correspond to active and dashed to passive. The rest of the columns contain direct comparisons between averaged observables computed for active and passive systems. Each figure features two $x$ axes: bottom is the P\'eclet number, $Pe$, in magenta and top is the temperature of the thermal bath, $Tk_B/\epsilon$, in green. The $y$-axis is shared between the two plots. Data corresponding to active systems (as functions of $Pe$) are plotted as magenta squares and data corresponding to passive systems (as functions of $Tk_B/\epsilon$) as green circles. The second column (\textbf{(b)}, \textbf{(f)} and \textbf{(j)}) contains the fraction of particles with ordered local environment: $P_{ico}$ for icosahedral symmetry in spherical, $\rho\sigma^3 = 0.155$ \textbf{(b)} and cylindrical clusters, $\rho\sigma^3 = 0.264$ \textbf{(f)} and $P_{hex}$ for hexagonal symmetry in lamellar clusters, $\rho\sigma^3 = 0.418$ \textbf{(j)}. The inset of \textbf{(j)} contains the number of icosahedral centers, $N_{ico}$. The third column (\textbf{(c)}, \textbf{(g)} and \textbf{(k)}) contains the average potential energy, $\langle u \rangle/\epsilon$ for the three densities under study. The inset of \textbf{(k)} shows a zoom in of $\langle u \rangle/\epsilon$ showing two hysteresis loops: the solid lines correspond to the branch of increasing $Pe$ and the dashed line to decreasing $Pe$. The fourth column (\textbf{(d)}, \textbf{(h)}) and \textbf{(l}) contains the effective diffusion coefficient, $D_{\small\text{eff}}\tau/\sigma^2$ for the three densities considered. The $y$-axis is on a logarithmic scale.}
    \label{fig:ActivevsPassive}
\end{figure*}

At equilibrium, for $\rho\sigma^3 = 0.155$, increasing the temperature leads first to a transition from a cluster-crystal to a cluster-fluid, which is accompanied by an internal rearranging of the particles forming the clusters that transit from a mostly local icosahedral symmetry to a disordered local environment. After that, the cluster-fluid undergoes a transition to a random percolating fluid. At $\rho\sigma^3 = 0.264$, as the temperature increases the system first transits from a hexagonal cylindrical phase with solid-like local icosahedral symmetry to a hexagonal cylindrical phase with a liquid-like local structure, and then to a percolating fluid. At $\rho\sigma^3 = 0.418$, upon increasing the temperature the system first undergoes a transition from lamellar phase with internal solid-like structure to a lamellar phase with internal liquid-like structure and then to a dense percolating fluid. All these transitions were previously described in reference \cite{serna2021formation}.

In the passive system, we can recover exactly the same transitions, at the three densities under study, thus confirming that even though in BD simulations dissipation is explicitly considered via the drag force in the translational equation of motion (see Equation \ref{e:BrownianTranslation}), the steady states reached at a given temperature of the thermal bath perfectly coincide with the equilibrium states at the same ensemble temperature in MD simulations. Unexpectedly, we find that gradually increasing the activity in systems previously equilibrated at low temperature, $Tk_B/\epsilon = 0.20$, induces transitions very similar to those observed in the passive system for the same three densities discussed above. In \textbf{Figures S1}, \textbf{S2} and \textbf{S3}, we provide snapshots and iso-probability surfaces of the positions PDF's for the typical phases observed in the active and passive systems.

Figure \ref{fig:ActivevsPassive} presents a complete summary of all the observables computed to characterize the structure and dynamics of the system for three selected densities at which the system self-assembles into cluster-crystal, hexagonal cylindrical and lamellar phases at a temperature at which the internal structure of the clusters (spherical clusters, cylinders and lamellae) remain highly ordered. Each row of Figure \ref{fig:ActivevsPassive} corresponds to one density, and each column to a given observable. The first column contains the pair correlation functions, $g(r)$, which provide the relative positioning of particles within the system at the particle-particle length scale, and it is typically acknowledged as a general structural property. The second column presents the fraction of particles with solid-like local environments. As reported in \cite{serna2021formation}, the local symmetry exhibited by spherical and cylindrical clusters is icosahedral whereas that of lamellar clusters is hexagonal. To quantify both types of symmetry we use the Common Neighbor Analysis (CNA) and the averaged six-fold bonding order parameter, $\bar{q}_6$ proposed by Lechner and Dellago \cite{lechner2008accurate}. The third column contains the average potential energy, $\langle u \rangle/\epsilon$ and the fourth column the effective diffusion coefficient, $D_{\small\text{eff}}\tau/\sigma^2$. Figures in the second, third and fourth columns contain two plots that share that same $y$-axis with different $x$-axis: the top one corresponds to the temperature of the thermal bath, $Tk_B/\epsilon$, and the bottom one to the P\'eclet number\cite{stenhammar2014phase}, $Pe = \frac{3v_a}{D_R\sigma}$ , that quantifies the level of activity relative to the noise present in the system. With such an arrangement of plots, we can fairly, and directly, compare the structure and dynamics of active and passive systems upon varying $Pe$ and $Tk_B/\epsilon$, respectively. The meaning of the colors used in Figure \ref{fig:ActivevsPassive} are explained in the caption. From now on, when we mention that two systems are structurally equivalent, we refer to them having the same, or very similar, global and local structural properties. With this criterion, for example, the active system at $Pe = 12.25$ and the passive system at $Tk_B/\epsilon = 0.35$ for $\rho\sigma^3 = 0.155$ are structurally equivalent systems because have roughly the same $g(r)$, $P_{ico}$ and $\langle u \rangle/\epsilon$, as can be seen from Figures \ref{fig:ActivevsPassive} a, b and c.

Let us start discussing the results for $\rho\sigma^3 = 0.155$. In Figure \ref{fig:ActivevsPassive}a, we observe that the three selected active/passive pairs exhibit quite similar $g(r)$, only differing appreciably for $r\geq2.5\sigma$. This indicates that the number of neighbors in the first two coordination shells is roughly the same in both cases. In Figure \ref{fig:ActivevsPassive}b, we present the fraction of particles with icosahedral environment, $P_{ico}$. As expected, increasing both the activity and the temperature decreases the icosahedral order. However, increasing $Tk_B/\epsilon$ seems to lower the local order faster than increasing $Pe$, for small values of $Pe$ and $Tk_B/\epsilon$. For $Tk_B/\epsilon \approx 0.30$ and $Pe \approx 8.50$ a cross-over of the two plots takes place at a $P_{ico} \approx 0.02$. At this temperature, the transition from a cluster-crystal to a cluster-fluid occurs, it being clearly signaled by a diffusive behavior at long times from the mean squared displacement (see \textbf{Figure S4}). However, as can be confirmed from \textbf{Figure S4}, the fully-developed long-time diffusive behavior is already observed at lower values of activity (for $Pe \approx 5.25$) for cases in which the local icosahedral order is still at high values compatible with a cluster-crystal (See Figure \ref{fig:ActivevsPassive}b), suggesting that activity promotes the formation of a new self-organized phase in which the cluster-fluid is made of clusters with a high local icosahedral order. From Figure \ref{fig:ActivevsPassive}c, where we present the average potential energy per particle, $\langle u\rangle/\epsilon$, we observe that the crossover takes place for $Tk_B/\epsilon = 0.35$ and $Pe = 12.25$, which correlates with the vanishing of the icosahedral local order in the active and passive systems (see Figure \ref{fig:ActivevsPassive}b). Although the potential energy is similar in the passive and active systems, at low $Pe$ the active system has lower potential energy compared to that of the passive system with equivalent structure.

Finally, in Figure \ref{fig:ActivevsPassive}d we report the effective diffusion coefficient, $D_{\small\text{eff}}\tau/\sigma^2$.Both in the passive and active systems the diffusion coefficient are only represented for those cases which exhibit a fully-developed diffusive behavior at long times (i.e. $Tk_B/\epsilon \geq 0.30$  and  $Pe\geq 5.25$, respectively). As discussed above, the pair $Pe = 5.25-Tk_B/\epsilon = 0.25$ shows a clear mismatch in the transport properties since for the active system we can extract a diffusion coefficient whereas in the passive case the system shows a sub-diffusive regime at long times. For the pair $Pe = 12.25-Tk_B/\epsilon = 0.35$,  $D_{\small\text{eff}}(Pe =12.25) = 0.25\sigma^2/\tau$  compared to the passive case, $D_{\small\text{eff}} (Tk_B/\epsilon=0.35) = 0.035 \sigma^2/\tau$ giving a diffusion ratio $D^a_{\small\text{eff}}/D^p_{\small\text{eff}} \approx 7.14$, with the superscripts meaning active ($a$) and passive ($p$). The diffusion ratio keeps growing with $Pe$, reaching a value $D^{a}_{eff}/D^{p}_{eff} \approx 12.4$, for the P\'eclet-temperature pair, $24.50-0.5$, which have essentially the same structure as seen from Figures \ref{fig:ActivevsPassive} a, b and c. Additional structural properties such as the cluster-cluster pair correlation functions, the cluster size distributions and the probability of finding a percolating state for both active and passive systems are presented in \textbf{Figure S5}, still showing a remarkable similarity.

Let us continue with the discussion of the results for $\rho\sigma^3 = 0.264$, at which the hexagonal cylindrical phase with internally ordered clusters is formed at low temperatures. As can be seen in Figure \ref{fig:ActivevsPassive}e, the low $Pe$ -low $Tk_B/\epsilon$ pair exhibits very similar radial distribution functions over all the studied distances.  This time, the low $Pe$- low $Tk_B/\epsilon$ pair exhibits approximately the same structure. For the other two pairs, slight discrepancies are observed for $r \geq 2.5\sigma$. From Figure \ref{fig:ActivevsPassive}f we can observe that the tendency is similar to that of the spherical clusters discussed earlier: the local icosahedral order is higher in the active system for low $Pe$ before a cross-over takes place around the pair $Pe = 5.25$-$Tk_B/\epsilon = 0.25$, from which the local icosahedral order decays faster than that of the passive one. Interestingly, the same transition in the internal structure of the cylinders from solid-like to liquid-like also occurs in the active system, signaled by the inflection point at $P_{ico} \approx 0.05$ for $Pe \approx 7.0$. As for the average potential energy, $\langle u \rangle/\epsilon$, presented in Figure \ref{fig:ActivevsPassive}g, passive and active systems exhibit almost the same values, with only very small differences at low $Pe$ and $Tk_B/\epsilon$, when the active system exhibits lower potential energy. Figure \ref{fig:ActivevsPassive}h contains the effective diffusion coefficient $D_{\small\text{eff}}\tau/\sigma^2$. Again, for the passive system, we can only extract the diffusivity for $Tk_B/\epsilon\geq0.30$, but this time the lowest $Pe$ required to achieve a fully-developed diffusive behavior in the active system is $Pe = 7.0$, which is a bit larger than that required for $\rho\sigma^3 = 0.155$ (see Figure \ref{fig:ActivevsPassive}d). Nevertheless, when comparing the ratio of diffusion coefficients at $Tk_B/\epsilon = 0.30$ and $Pe = 7.0$, $D^{a}_{eff}/D^{p}_{eff} \approx 1.72$, meaning that for the cylindrical phase the decoupling between structure and dynamics occurs at lower $Pe$. The ratio again grows, adopting the value $D^{a}_{eff}/D^{p}_{eff} \approx 10.42$ at the highest temperature and P\'eclet which have equivalent global and local structure and potential energy (see Figures \ref{fig:ActivevsPassive} e,f and g).

We now discuss the results obtained for $\rho\sigma^3 = 0.418$. The $g(r)$, shown in Figure \ref{fig:ActivevsPassive}i, is again almost identical for the passive and active systems for the selected P\'eclet-temperature pairs. However, comparison of the fraction of particles with hexagonal local environment (Figure \ref{fig:ActivevsPassive}j) reveals important structural differences, as temperature/activity increases. Whereas at low temperature and activity almost all particles have solid-like local environments, the transition to liquid-like local environments seems somewhat steeper in the active than in the passive. In reference \cite{serna2021formation}, it was shown that upon gradual heating the transition from lamellae with internal solid-like structure to lamellae with internal liquid-like structure occurs preceded by a slight decrease in $P_{hex}$ similarly to what we observe here in the passive case at $Tk_B/\epsilon = 0.30$, with the only difference that this effect is amplified in the latter case, probably due to finite size effects experienced by the smaller system studied in \cite{serna2021formation}. Interestingly, when cooling the system, the transition appears steeper than that observed during heating, as discussed in  \cite{serna2021formation} and further confirmed here by additional heating and cooling simulations of the passive system (not shown). To ensure that the structure observed at $Tk_B/\epsilon = 0.30$ corresponds to the steady state, we run long simulations for $1.32\times10^8$ time steps for two independent systems, obtaining the same results. The reason behind this partial drop of $P_{hex}$ is that some portions of the lamellae start to melt while the majority of them remain in the internal solid-like state, something that in the smaller system was barely noticed but prominent in the larger one considered here. Our preliminary observations suggest that the liquid-like portions of the lamellae start to develop from the incomplete third layer of particles that the lamellae have. These findings provide some insights into the nature of the intra-cluster transition of the lamellar phase at equilibrium and its kinetics. However, it is beyond the scope of the present article, and a more detailed analysis of this is left for future research. Coming back to the results of the active system, we can say now that increasing $Pe$ modifies the way the lamellar phase transitions from a solid-like internal structure to a liquid-like internal structure, making it steeper, probably due to modifications of the kinetics of the transition. Again, we leave this analysis for future work.


Interestingly, this transition is also signaled by sudden increase of particles with icosahedral order (see inset of Figure \ref{fig:ActivevsPassive}j). The increase of the icosahedral order is three times greater in the active case than in the passive case, revealing that increasing the activity betters the local icosahedral order at the first neighbors length-scale. Now, looking at $\langle u \rangle / \epsilon$ in Figure \ref{fig:ActivevsPassive}k, we observe that at this higher density ($\rho\sigma^3 = 0.418$), the potential energy shows the largest discrepancies between the active and passive systems. For $Tk_B/\epsilon \leq 0.30$ and $Pe \leq 7.0$, both the passive and active systems mostly form solid-like lamellae, however the potential energy in the active system is significantly lower than that of the passive, pointing out a most compact and energetically stable structure promoted by low activity. The lamellae become liquid at $Tk_B/\epsilon \approx 0.35$ and $Pe \approx 8.25$ in the passive and active systems, respectively. For both systems, the transition is signaled by a jump in the potential energy, being the steeper in the active case. The transition from liquid lamellae to dense percolating fluid occurs at $Tk_B/\epsilon \approx 0.45$ and $Pe \approx 9.50$, again signaled by a sudden increase in the potential energy in both systems. Coming back to the emergence of icosahedral order presented in the inset of Figure \ref{fig:ActivevsPassive}j, we notice that in the passive system it happens only when lamellae are liquid to then disappear once the dense percolating fluid is formed. In contrast, in the active system, the icosahedral order prevails even when the lamellae melt and only vanishes for $Pe > 18$ when the system is deep within the stability region of the dense percolating fluid. 
This further confirms that higher local order is preserved in active systems even at high activities.   

At equilibrium, these transitions are accompanied by two hysteresis loops when gradually heating and cooling the systems as shown in \cite{serna2021formation}, suggesting the first-order nature of the transitions. Surprisingly, this hysteretic behavior also applies for the active system when gradually increasing and decreasing $Pe$, as shown in the inset of Figure \ref{fig:ActivevsPassive}k. The general structural and dynamical features of the observed phases in the increasing and decreasing $Pe$ runs are essentially the same. This is not generally true for active systems where different non-equilibrium pathways or different initial conditions might lead to different steady states even at the same set of parameters \cite{chen2023initial, huber2018emergence, prymidis2015self, dabelow2019irreversibility,dabelow2025thermodynamic,kuan2015hysteresis}. In the active system, all the transitions occur at low to intermediate $Pe$ so the system remains relatively close to equilibrium, and given the fact that the interparticle interactions lead to well-defined structures at low temperatures, the interplay of these two contributions might explain the hysteresis loops upon changes in the activity. Hysteresis loops have been previously observed at the onset of MIPS \cite{solon2015active} and jamming transitions \cite{yang2022interplay}. A more detailed understanding of this phenomenon is left for future work. To finalize with this section, we present $D_{\small\text{eff}}\tau/\sigma^2$ in Figure \ref{fig:ActivevsPassive}l. The mean squared displacements are available in \textbf{Figure S4}. This time, the structure-dynamics decoupling is clearly evident for the lowest P\'eclet-temperature pair. The principle is the same: very similar global and local structure with a decoupled diffusion coefficient, starting with a ratio $D^{a}_{eff}/D^{p}_{eff} \approx 9.60$ for $Pe = 8.75$ and $Tk_B/\epsilon = 0.30$ and reaching $D^{a}_{eff}/D^{p}_{eff} \approx 7.62$, thus constituting the only density of those considered here for which the ratio decreases with $Pe$.

Now, we can say that for the three ordered microphases considered, there exist a pair $Tk_B/\epsilon$-$Pe$ from which a structure-dynamics decoupling occurs. At equilibrium, structure and dynamics of suspensions of hard spheres have been shown to be closely related and thus one might be inferred from the other \cite{fuchs1992primary, fuchs1999aspects,marin2019slowing,marin2020tetrahedrality,boattini2020autonomously}. On the other hand, soft colloids, might exhibit a decoupling between structural and dynamical properties mediated by interactions as has been recently shown in \cite{arenas2025structure}. In particular, SALR systems are known to exhibit dynamical arrest under different conditions, which can be also acknowledged as a sort of structure-dynamics decoupling \cite{royall2015role}. In this case, the transport properties of the SALR fluids are significantly hindered even though their structural properties suggest states with faster dynamics. The mechanisms behind dynamical arrest in SALR fluids are usually related with frustration of the microphase separation, and clusters and percolating networks formation that lead to different non-ergodic states such as gels and cluster glasses\cite{toledano2009colloidal, klix2010structural,campbell2005dynamical,ruiz2021role,cardinaux2011cluster,bomont2024arrested}. In this article, we demonstrate how non-equilibrium pathways lead to structure-dynamics decoupling that instead enhance the diffusion while keeping the average structural properties, in essence, unaltered.

To finish this section, we want to mention that the structure of the percolating fluids formed at high $Pe$ for $\rho\sigma^3 = 0.264$ and $\rho\sigma^3 = 0.418$ resemble a disordered bicontinuous phase that evolve in time. Recently, continuously reconfiguring bicontinuous phases were observed in experimental mixtures of active-passive fluids \cite{gulati2026dynamical} with a notable similarity to those observed in this article (see \textbf{Figure S6}). Although bicontinuous phases are out of the scope of the article, the model described here might be used as a theoretical and simulation platform to study this type of phases in active matter.

\subsection{Effective pressure and temperature}

So far, we have characterized the structure-dynamics decoupling induced by activity. We identify such a decoupling as the unmatched transport properties of active and passive systems given an equivalent (or very similar) local and global structure. Thus, we can clearly distinguish the transitions induced through equilibrium and non-equilibrium pathways just by looking at the transport properties. However, when considering the structure of the system solely, we might be tempted to relate the effects of the activity as a mere increase of the effective pressure and temperature of the system, especially if we acknowledge the reversible nature of the transitions induced by varying $Pe$ as discussed before (see Figure \ref{fig:ActivevsPassive}k). With the purpose of gaining more insights on the effects of activity on the thermodynamic properties of the system, we calculate the effective pressure, $p_{_\text{eff}}\sigma^3/\epsilon$ and the configurational temperature, $T_{_\text{conf}}k_B/\epsilon$, following Equations \ref{e:p_eff} and \ref{e:T_conf}, respectively.

\begin{figure}[ht!]
    \centering    \includegraphics[width=0.25\textwidth]{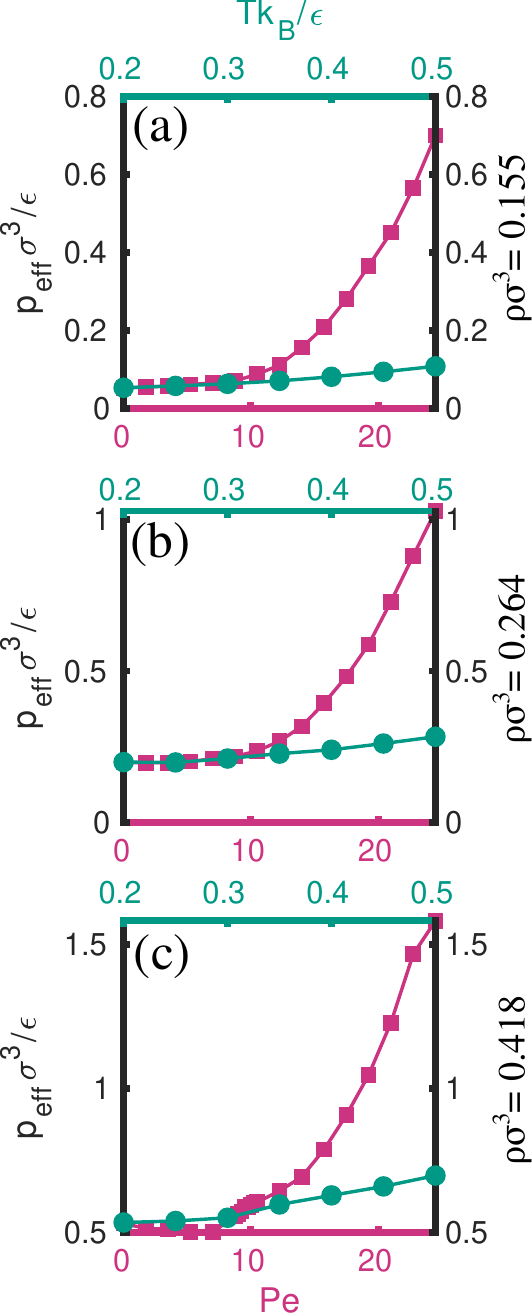}
    \caption{\textit{Comparing the effective pressure in the active and passive system}. The effective pressure, $p_{\text{eff}}\sigma^3/\epsilon$, as a function of the P\'eclet number, $Pe$, and the temperature, $Tk_B/\epsilon$, for the three densities considered, \textbf{(a)} $\rho\sigma^3 = 0.155$, \textbf{(b)} $\rho\sigma^3 = 0.264$, and \textbf{(c)} $\rho\sigma^3 = 0.418$.}
    \label{fig:Pressure}
\end{figure}

In Figure \ref{fig:Pressure}, we present the comparison of $p_{_\text{eff}}\sigma^3/\epsilon$ between the active and passive systems. For $\rho\sigma^3 = 0.155$ (see Figure \ref{fig:Pressure}a) and $\rho\sigma^3 = 0.264$ (\ref{fig:Pressure}b), the effective pressure in the active system remains very close to that of the passive counterpart at low P\'eclet ($Pe \leq 12$), indicating that the system remains close to the equilibrium steady state. For $Pe > 12$, the effective pressure of the active systems starts to grow monotonically as $Pe$ does. Remarkably, this change in the tendencies of $p_{_\text{eff}}\sigma^3/\epsilon$ in the active and passive systems fairly concurs with the structure-dynamics decoupling discussed in the previous section, implying that such a decoupling requires the system being sufficiently far from equilibrium to happen although the structural properties remain roughly the same (see the first two rows of Figure \ref{fig:ActivevsPassive}). For $\rho\sigma^3 = 0.418$ (see Figure \ref{fig:Pressure}c), the separation of the $p_{_\text{eff}}\sigma^3/\epsilon$ curves in the active and passive systems also occurs around $Pe\approx12$, again corresponding to the onset of the structure-dynamics decoupling in the lamellar phase (see Figure \ref{fig:ActivevsPassive}l). However, unlike the other densities, for low activities $Pe < 12$, the effective pressure of the active system is considerably smaller than that of its passive counterpart, correlating with the lower potential energy at this low activity regime already discussed (see Figure \ref{fig:ActivevsPassive}k). Thus, in the solid-like lamellar phase, low activity reconfigures the local environment of the particles and thus the system achieves more energetically stable configurations and lower effective pressures.

\begin{figure}[htb!]
    \centering    \includegraphics[width=0.30\textwidth]{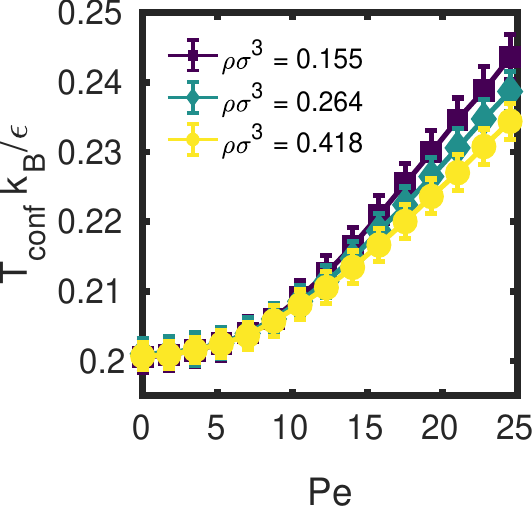}
    \caption{The configurational temperature, $T_\text{conf}k_B/\epsilon$, as a function of $Pe$ for the three densities considered }
    \label{fig:Temperature}
\end{figure}

Coming back to the onset of the structure-dynamics decoupling at $Pe \approx 12$ for all the ordered microphases considered, it should be noticed that $Pe = 12$ corresponds to $F_a = 2.40\epsilon/\sigma$, which is close to the force minimum of the interaction potential in the attractive well ($\approx3.0\epsilon/\sigma$, see the inset of the top panel of Figure \ref{fig:SALR}), so the onset of the decoupling rather matches with the active force that allows particles to escape from clusters, especially those on their surface.       

Although a unified and conceptually robust definition of temperature for non-equilibrium systems is still missing, some definitions inherited from equilibrium thermodynamics are useful in the context of active matter. In the recent article by Hecht et al. \cite{hecht2024define}, the authors discuss different alternatives to measure temperature in active systems, pointing out kinetic definitions, based on the equipartition theorem, and configurational definitions as the most reliable. One might think that since active matter systems are usually modeled in the overdamped regime, the best temperature definition should be that derived from the fluctuation-dissipation theorem, $T_{ein}k_B = \gamma D_{\small\text{eff}}$, which has been used in  different types of active systems and extended for conditions far from equilibrium \cite{berthier2017active,cugliandolo2019effective,burkholder2019fluctuation,caprini2021fluctuation,boriskovsky2024fluctuation}. However, in its simplest version, it is not a good choice for systems with attractive interactions or high densities which very often exhibit sub-diffusion \cite{toledano2009colloidal, prymidis2015self}. Note that since the friction coefficient, $\gamma = 1.0m/\tau$, 
the value of $D_{\small\text{eff}}\tau/\sigma^2$ coincides with that of $T_{ein}k_B/\epsilon$ from the Stokes-Einstein relation $D = k_BT/\gamma$ \cite{hecht2024define}, and we can directly see from the third column of Figure  \ref{fig:ActivevsPassive}, that it is far from coincident with $Tk_B/\epsilon$ even in the passive system due to the complex SALR interactions. Nonetheless, the configurational temperature calculated from Equation \ref{e:T_conf}, perfectly coincides with $Tk_B/\epsilon$ in the passive system and is independent of $\rho\sigma^3$ as can be seen from \textbf{Figure S7}. Hence, $T_{conf}$ not only provides a consistent definition that recovers the temperature of thermal bath at $Pe = 0$, but also contains structural and energetic information that is very useful for understanding the contributions of activity to the decoupling between dynamics and structure.

Thus, we report in Figure \ref{fig:Temperature}, the configurational temperature of the system at the three selected densities. $T_\text{conf}k_B/\epsilon$ shows an expected increasing tendency for all densities. Although as $Pe$ increases the differences between the three curves become more noticeable, such differences are still within the error and, at least for the range of $Pe$ studied in this article, we can say that $T_\text{conf}$ is essentially independent of density in the active system which is also required for a good definition of a non-equilibrium temperature. In contrast, the fluctuation-dissipation based definition is strongly dependent on density for the same range of $Pe$ as can be seen from Figures \ref{fig:ActivevsPassive} d, h and l. At higher activities, $T_\text{conf}$ still exhibits a fair density independence as can be seen from \textbf{Figure S8}. We should emphasize that our intention with comparing different temperature definitions is two fold: first, to highlight positive and negative aspects of definitions inherited from equilibrium thermodynamics, thus contributing to the debate in the field when complex interactions are at play; second, to choose the most consistent definition for the particular system we consider here.

Intriguingly, the configurational temperature changes at a significantly lower rate than the other structural properties measured (see Figure \ref{fig:ActivevsPassive}), reaching values of $T_\text{conf}k_B/\epsilon \approx 0.24$ for the three densities when the systems reach steady states compatible with a random percolating fluid ($\rho\sigma^3 = 0.155$), a percolating fluid ($\rho\sigma^3 = 0.264$) and a dense percolating fluid ($\rho\sigma^3 = 0.418$), phases that also appear in passive systems but st significantly higher temperatures as already discussed (see also \textbf{Figures S1}, \textbf{S2} and \textbf{S3}). So, increasing the activity modifies the global structure, but maintaining to some extent the local order within the clusters. The $T_\text{conf}$ perfectly reflects this since, given the definition in Equation \ref{e:T_conf}, and the truncated and shifted nature of the potential in Equation \ref{e:USALR}, only pairs of particles with relative distances below $r_c = 4.0\sigma$ will contribute to $T_\text{conf}$. In other words, although increasing the activity favors the mobility of particles, the intra-cluster structure is kept similar to that imposed by the thermal bath $Tk_B/\epsilon$ in the passive system. It is like the active system retains some memory of the local structure at the temperature of the thermal bath. This phenomenon can be clearly recognized in the cluster-crystal and the lamellar phases. In the former, we have the formation of a cluster-fluid made out of clusters with icosahedral symmetry and in the latter the highest energetic stability of the active solid-like lamellae at low $Pe$ (see Figure \ref{fig:ActivevsPassive}k) and increase of the icosahedral order in a range of $Pe$ where the liquid-like lamellae and the dense percolating fluid are stable (see Figure \ref{fig:ActivevsPassive}j). Lastly, \textbf{Figure S9} of the supplementary material displays the non-equilibrium equation of state as $p_{_\text{eff}}\sigma^3/\epsilon$ vs $T_\text{conf}k_B/\epsilon$.

\section{\label{sec:Conlcusions}Conclusions}

We have performed extensive BD simulations of active colloidal particles whose effective interactions feature short-ranged attraction, coming from depletion, and long-ranged repulsion, coming from screened electrostatics (SALR). The particles undergo overdamped translational and rotational dynamics. We investigated the structural and dynamical properties of the system upon changes in the activity. Surprisingly, we found that, starting from a totally equilibrated passive system at low temperature for three selected densities, increasing the self-propulsion force leads to transitions that resemble, from the structural perspective, those observed in the passive system when the temperature is increased (see Figure \ref{fig:ActivevsPassive}). The structural resemblance is such, that at first sight the effect of activity on the SALR particles suspension seems to be a mere effective increase in the pressure and the temperature. This first impression is reinforced by the fact that upon gradual increasing and subsequent decreasing of the activity, the active system exhibits hysteresis and reversibility in the phase behavior (see Figure \ref{fig:ActivevsPassive} k), something that usually happens in passive and equilibrium systems but does not happen in out-of-equilibrium systems in general.  

However, when looking at the diffusion coefficient of the active system, we observe that beyond a certain value of the activity, it increases much faster than the diffusion coefficients typically obtained in the passive system as the temperature increases, whereas the structure remains strikingly similar between the active and passive systems. We refer to this phenomenon as an activity-induced structure-dynamics decoupling. This type of decoupling has been already characterized in passive soft colloids and in equilibrium systems with competing interactions as dynamically arrested states. In the latter case, the decoupling usually occurs the other way around: two independent systems at the same conditions exhibit the same sub-diffusive behavior but different structures, usually because one of them has not reached the equilibrium state due to a slow dynamics \cite{toledano2009colloidal, klix2010structural}.

We have further characterized the behavior of the active system by calculating the effective pressure and the configurational temperature. We observe that the onset of the structure-dynamics decoupling fairly coincides with the separation of the effective pressure curves for active and passive systems (see Figure \ref{fig:Pressure}), indicating that the activity-induced structure-dynamics decoupling only happens for system sufficiently out-of-equilibrium. Although the correlation is not perfect, the onset of the decoupling rather corresponds to the active particles having a self-propulsion force close to the force minimum of the SALR potential, thus promoting the exchange of particles between clusters and enhancing the diffusion. On the other hand, the configurational temperature only increases slightly as the activity increases, even when the diffusion coefficient of the active system is higher than that of the passive system (see Figures \ref{fig:ActivevsPassive} d h and l, and Figure \ref{fig:Temperature}). This suggests that the interactions modulate the dynamic behavior of active systems and induces a sort of structural memory on the system that, even at sufficiently high activities at which global structural transitions occur and the mobility of particles is enhanced, retains local structural characteristics of the corresponding equilibrium structure at the temperature of the thermal bath. This memory effect is clearly observed in the cluster-fluid made out of locally ordered clusters with icosahedral symmetry described in this article (see Figures \ref{fig:ActivevsPassive} b, c, d), and in the enhanced icosahedral order exhibited by the active system when the solid-like lamellar phase becomes a liquid-like lamellar phase and even in the dense percolating phase (see Figure \ref{fig:ActivevsPassive}j).

We observe the formation of bicontinuous phases, although not ordered, in the percolating fluids obtained at intermediate activities for $\rho\sigma^3 = 0.264$ and $0.418$ similar to those obtained in a recent work combining theory and experiments of mixtures of active and passive fluids \cite{gulati2026dynamical}. The model considered here can be a platform for numerical studies on this type of phases. On the other hand, in a future work we plan to address the problem of the nature and kinetics of the intra-cluster transition in the lamellar phase and the fundamental differences when achieved via equilibrium and non-equilibrium pathways.

The findings of this study highlight the fundamental role of SALR interactions in the emergence of structural memory within active matter systems. In these active SALR assemblies, the underlying ordered microphase architecture remains robust despite the strong enhancement of transport within the system induced by activity, revealing a pronounced decoupling between structure and dynamics. This persistence of structural organization under intrinsically out-of-equilibrium conditions provides a simple yet powerful mechanistic framework for understanding the stability and dynamics of biocondensates and membraneless organelles inside living cells  \cite{sweatman2019giant,bores2026association}. In particular, the emergent structural memory in active cytoeskeletal composites has been recently highlighted as fundamental in the consistency of intracellular organization despite its permanent renewing \cite{kuvcera2022actin}. Thus, active SALR systems might be used to explore, theoretically and numerically, the implications of structural memory and its link with functionality in biological systems. More broadly, these results demonstrate that active systems with competing interactions offer a promising route toward the design of smart metamaterials capable of combining enhanced transport properties with long-lived structural memory and reversible self-organization. From a fundamental perspective, the results discussed here might contribute to theoretical developments of robust temperature definitions in out-of-equilibrium systems where interactions and structural features mediate the dynamics beyond collisions and volume exclusion.


\section{\label{sec:SI}Supplementary Information}

The supplementary information contains additional Figures and discussions \textbf{Figures S1, S2, S3:} Configurations and iso-probability surfaces for both active and passive systems. \textbf{Figure S4:} Mean squared displacement of active and passive systems. \textbf{Figure S5:} Additional structural observables for the active and passive systems at $\rho\sigma^3 = 0.154$. \textbf{Figure S6:} Instantaneous iso-density surfaces for the active percolating fluids. \textbf{Figure S7:} Configurational temperature of the passive system. \textbf{Figure S8:} Configurational temperature and effective temperature obtained from fluctuation-dissipation theorem at high $Pe$. \textbf{Figure S9:} The non-equilibrium equation of state $p_{eff}\sigma^3/\epsilon$ vs $T_{conf}k_B/\epsilon.$

\begin{acknowledgments}
H.S acknowledges funding from the Marie Skłodowska-Curie Grant Agreement No. 101108868 (BIOMICAR) for the acquisition of computational facilities that were used to carry out this work. J.M.R. acknowledges funding from Juan de la Cierva postdoctoral fellowship  funding from Spanish Ministry of Education (JDC2024-053228-I). E.G.N. acknowledges financial support with grant PID2023-151751NB-I00 funded by MCIN/AEI / 10.13039/501100011033. 
\end{acknowledgments}

\bibliography{apssamp}

@article{bores2026association,
  title={Association and phase transitions in simple models for biological and soft matter condensates},
  author={Bores, Cecilia and Diaz-Pozuelo, Antonio and Lomba, Enrique},
  journal={The Journal of Chemical Physics},
  volume={164},
  number={5},
  year={2026},
  publisher={AIP Publishing}
}

@article{sweatman2019giant,
  title={The giant SALR cluster fluid: A review},
  author={Sweatman, Martin B and Lue, Leo},
  journal={Advanced Theory and Simulations},
  volume={2},
  number={7},
  pages={1900025},
  year={2019},
  publisher={Wiley Online Library}
}

@article{LAMMPS,
  author = "A. P. Thompson and H. M. Aktulga and R. Berger and 
     D. S. Bolintineanu and W. M. Brown and P. S. Crozier and
     P. J. in 't Veld and A. Kohlmeyer and S. G. Moore and T. D. Nguyen and
     R. Shan and M. J. Stevens and J. Tranchida and C. Trott and S. J. Plimpton",
  title = "{LAMMPS} - a flexible simulation tool for
     particle-based materials modeling at the 
     atomic, meso, and continuum scales",
  journal = "Comp. Phys. Comm.",
  volume =  "271",
  pages =   "108171",
  year =    "2022",
  doi = "10.1016/j.cpc.2021.108171"
}

@article{chacon2022intrinsic,
  title={Intrinsic structure perspective for MIPS interfaces in two-dimensional systems of active Brownian particles},
  author={Chac{\'o}n, Enrique and Alarc{\'o}n, Francisco and Ram{\'\i}rez, Jorge and Tarazona, Pedro and Valeriani, Chantal},
  journal={Soft Matter},
  volume={18},
  number={13},
  pages={2646--2653},
  year={2022},
  publisher={Royal Society of Chemistry}
}

@article{serna2021formation,
  title={Formation and internal ordering of periodic microphases in colloidal models with competing interactions},
  author={Serna, Horacio and Pozuelo, Antonio D{\'\i}az and Noya, Eva G and G{\'o}{\'z}d{\'z}, Wojciech T},
  journal={Soft Matter},
  volume={17},
  number={19},
  pages={4957--4968},
  year={2021},
  publisher={Royal Society of Chemistry}
}

@incollection{gentile2023active, 
   author = {Gentile, Luigi and Kurzthaler, Christina and Stone, Howard A.}, 
   isbn = {978-1-83916-229-9}, 
   title = {What is ‘Active Matter’??}, 
   booktitle = {Out-of-equilibrium Soft Matter}, 
   publisher = {The Royal Society of Chemistry}, 
   year = {2023}, 
   month = {03}, 
   doi = {10.1039/9781839169465-00001}, 
   url = {https://doi.org/10.1039/9781839169465-00001}, 
}

@article{ramaswamy2017active,
doi = {10.1088/1742-5468/aa6bc5},
url = {https://doi.org/10.1088/1742-5468/aa6bc5},
year = {2017},
month = {may},
publisher = {IOP Publishing and SISSA},
volume = {2017},
number = {5},
pages = {054002},
author = {Ramaswamy, Sriram},
title = {Active matter},
journal = {Journal of Statistical Mechanics: Theory and Experiment}
}

@article{vicsek2012collective,
  title={Collective motion},
  author={Vicsek, Tam{\'a}s and Zafeiris, Anna},
  journal={Physics reports},
  volume={517},
  number={3-4},
  pages={71--140},
  year={2012},
  publisher={Elsevier}
}

@article{cates2015motility,
  title={Motility-induced phase separation},
  author={Cates, Michael E and Tailleur, Julien},
  journal={Annu. Rev. Condens. Matter Phys.},
  volume={6},
  number={1},
  pages={219--244},
  year={2015},
  publisher={Annual Reviews}
}

@article{omar2023mechanical,
  title={Mechanical theory of nonequilibrium coexistence and motility-induced phase separation},
  author={Omar, Ahmad K and Row, Hyeongjoo and Mallory, Stewart A and Brady, John F},
  journal={Proceedings of the National Academy of Sciences},
  volume={120},
  number={18},
  pages={e2219900120},
  year={2023},
  publisher={National Academy of Sciences}
}

@article{martin2021characterization,
  title={Characterization of MIPS in a suspension of repulsive active Brownian particles through dynamical features},
  author={Martin-Roca, Jos{\'e} and Martinez, Raul and Alexander, Lachlan C and Diez, Angel Luis and Aarts, Dirk GAL and Alarcon, Francisco and Ram{\'\i}rez, Jorge and Valeriani, Chantal},
  journal={The Journal of Chemical Physics},
  volume={154},
  number={16},
  year={2021},
  publisher={AIP Publishing}
}

@article{paoluzzi2022motility,
  title={From motility-induced phase-separation to glassiness in dense active matter},
  author={Paoluzzi, Matteo and Levis, Demian and Pagonabarraga, Ignacio},
  journal={Communications Physics},
  volume={5},
  number={1},
  pages={111},
  year={2022},
  publisher={Nature Publishing Group UK London}
}

@article{stenhammar2014phase,
  title={Phase behaviour of active Brownian particles: the role of dimensionality},
  author={Stenhammar, Joakim and Marenduzzo, Davide and Allen, Rosalind J and Cates, Michael E},
  journal={Soft matter},
  volume={10},
  number={10},
  pages={1489--1499},
  year={2014},
  publisher={Royal Society of Chemistry}
}

@article{liu2019self,
  title={Self-driven phase transitions drive Myxococcus xanthus fruiting body formation},
  author={Liu, Guannan and Patch, Adam and Bahar, Fatmag{\"u}l and Yllanes, David and Welch, Roy D and Marchetti, M Cristina and Thutupalli, Shashi and Shaevitz, Joshua W},
  journal={Physical review letters},
  volume={122},
  number={24},
  pages={248102},
  year={2019},
  publisher={APS}
}

@article{van2019interrupted,
  title={Interrupted motility induced phase separation in aligning active colloids},
  author={Van Der Linden, Marjolein N and Alexander, Lachlan C and Aarts, Dirk GAL and Dauchot, Olivier},
  journal={Physical review letters},
  volume={123},
  number={9},
  pages={098001},
  year={2019},
  publisher={APS}
}

@article{buttiononi2013dynamical,
  title = {Dynamical Clustering and Phase Separation in Suspensions of Self-Propelled Colloidal Particles},
  author = {Buttinoni, Ivo and Bialk\'e, Julian and K\"ummel, Felix and L\"owen, Hartmut and Bechinger, Clemens and Speck, Thomas},
  journal = {Phys. Rev. Lett.},
  volume = {110},
  issue = {23},
  pages = {238301},
  numpages = {5},
  year = {2013},
  month = {Jun},
  publisher = {American Physical Society},
  doi = {10.1103/PhysRevLett.110.238301},
  url = {https://link.aps.org/doi/10.1103/PhysRevLett.110.238301}
}

@article{bates1990block,
  title={Block copolymer thermodynamics: theory and experiment},
  author={Bates, Frank S and Fredrickson, Glenn H},
  journal={Annual review of physical chemistry},
  volume={41},
  number={1},
  pages={525--557},
  year={1990}
}

@article{alexandridis1998record,
  title={A record nine different phases (four cubic, two hexagonal, and one lamellar lyotropic liquid crystalline and two micellar solutions) in a ternary isothermal system of an amphiphilic block copolymer and selective solvents (water and oil)},
  author={Alexandridis, Paschalis and Olsson, Ulf and Lindman, Bj{\"o}rn},
  journal={Langmuir},
  volume={14},
  number={10},
  pages={2627--2638},
  year={1998},
  publisher={ACS Publications}
}

@article{yang2022interplay,
  title={Interplay between jamming and motility-induced phase separation in persistent self-propelling particles},
  author={Yang, Jing and Ni, Ran and Ciamarra, Massimo Pica},
  journal={Physical Review E},
  volume={106},
  number={1},
  pages={L012601},
  year={2022},
  publisher={APS}
}

@article{solon2015active,
  title={Active brownian particles and run-and-tumble particles: A comparative study},
  author={Solon, Alexandre and Cates, M and Tailleur, J},
  journal={The European Physical Journal. Special Topics},
  volume={224},
  number={7},
  pages={1231--1262},
  year={2015}
}

@article{mognetti2013living,
  title={Living clusters and crystals from low-density suspensions of active colloids},
  author={Mognetti, Bortolo Matteo and {\v{S}}ari{\'c}, An{\dj}ela and Angioletti-Uberti, Stefano and Cacciuto, A and Valeriani, C and Frenkel, Daan},
  journal={Physical review letters},
  volume={111},
  number={24},
  pages={245702},
  year={2013},
  publisher={APS}
}

@article{prymidis2015self,
  title={Self-assembly of active attractive spheres},
  author={Prymidis, Vasileios and Sielcken, Harmen and Filion, Laura},
  journal={Soft Matter},
  volume={11},
  number={21},
  pages={4158--4166},
  year={2015},
  publisher={Royal Society of Chemistry}
}

@article{tung2016micro,
  title={Micro-phase separation in two dimensional suspensions of self-propelled spheres and dumbbells},
  author={Tung, Clarion and Harder, Joseph and Valeriani, Chantal and Cacciuto, Angelo},
  journal={Soft Matter},
  volume={12},
  number={2},
  pages={555--561},
  year={2016},
  publisher={Royal Society of Chemistry}
}

@article{cao2024conceptual,
  title={A conceptual framework to understand the self-assembly of chemically active colloids},
  author={Cao, Dezhou and Yan, Zuyao and Cui, Donghao and Khan, Mohd Yasir and Duan, Shifang and Xie, Guoqiang and He, Zikai and Xing, Ding Yu and Wang, Wei},
  journal={Langmuir},
  volume={40},
  number={21},
  pages={10884--10894},
  year={2024},
  publisher={ACS Publications}
}

@article{imperio2006microphase,
  title={Microphase separation in two-dimensional systems with competing interactions},
  author={Imperio, A and Reatto, L},
  journal={The Journal of chemical physics},
  volume={124},
  number={16},
  year={2006},
  publisher={AIP Publishing}
}

@article{ciach2013origin,
  title={Origin of similarity of phase diagrams in amphiphilic and colloidal systems with competing interactions},
  author={Ciach, A and P{\k{e}}kalski, J and G{\'o}{\'z}d{\'z}, WT},
  journal={Soft Matter},
  volume={9},
  number={27},
  pages={6301--6308},
  year={2013},
  publisher={Royal Society of Chemistry}
}

@article{pini2017pattern,
  title={Pattern formation and self-assembly driven by competing interactions},
  author={Pini, Davide and Parola, Alberto},
  journal={Soft Matter},
  volume={13},
  number={48},
  pages={9259--9272},
  year={2017},
  publisher={Royal Society of Chemistry}
}

@article{zhuang2016equilibrium,
  title={Equilibrium phase behavior of a continuous-space microphase former},
  author={Zhuang, Yuan and Zhang, Kai and Charbonneau, Patrick},
  journal={Physical review letters},
  volume={116},
  number={9},
  pages={098301},
  year={2016},
  publisher={APS}
}

@article{al2022dual,
  title={Dual nature of magnetic nanoparticle dispersions enables control over short-range attraction and long-range repulsion interactions},
  author={Al Harraq, Ahmed and Hymel, Aubry A and Lin, Emily and Truskett, Thomas M and Bharti, Bhuvnesh},
  journal={Communications Chemistry},
  volume={5},
  number={1},
  pages={72},
  year={2022},
  publisher={Nature Publishing Group UK London}
}

@article{stukowski2010visualization,
  title={Visualization and analysis of atomistic simulation data with OVITO--the Open Visualization Tool},
  author={Stukowski, Alexander},
  journal={Modelling and simulation in materials science and engineering},
  volume={18},
  number={1},
  pages={015012},
  year={2010}
}

@article{lechner2008accurate,
  title={Accurate determination of crystal structures based on averaged local bond order parameters},
  author={Lechner, Wolfgang and Dellago, Christoph},
  journal={The Journal of chemical physics},
  volume={129},
  number={11},
  year={2008},
  publisher={AIP Publishing}
}

@article{faken1994systematic,
  title={Systematic analysis of local atomic structure combined with 3D computer graphics},
  author={Faken, Daniel and J{\'o}nsson, Hannes},
  journal={Computational Materials Science},
  volume={2},
  number={2},
  pages={279--286},
  year={1994},
  publisher={Elsevier}
}

@article{chen2023initial,
  title={Initial-state dependence of phase behaviors in a dense active system},
  author={Chen, Lu and Zhang, Bokai and Tu, ZC},
  journal={Chinese Physics B},
  volume={32},
  number={8},
  pages={086401},
  year={2023},
  publisher={Chinese Physical Society and IOP Publishing Ltd}
}

@article{huber2018emergence,
  title={Emergence of coexisting ordered states in active matter systems},
  author={Huber, L and Suzuki, R and Kr{\"u}ger, T and Frey, E and Bausch, AR},
  journal={Science},
  volume={361},
  number={6399},
  pages={255--258},
  year={2018},
  publisher={American Association for the Advancement of Science}
}

@article{dabelow2019irreversibility,
  title={Irreversibility in active matter systems: Fluctuation theorem and mutual information},
  author={Dabelow, Lennart and Bo, Stefano and Eichhorn, Ralf},
  journal={Physical Review X},
  volume={9},
  number={2},
  pages={021009},
  year={2019},
  publisher={APS}
}

@article{dabelow2025thermodynamic,
  title={Thermodynamic nature of irreversibility in active matter},
  author={Dabelow, Lennart and Eichhorn, Ralf},
  journal={Physical Review Research},
  volume={7},
  number={3},
  pages={033077},
  year={2025},
  publisher={APS}
}

@article{kuan2015hysteresis,
  title={Hysteresis, reentrance, and glassy dynamics in systems of self-propelled rods},
  author={Kuan, Hui-Shun and Blackwell, Robert and Hough, Loren E and Glaser, Matthew A and Betterton, MD},
  journal={Physical Review E},
  volume={92},
  number={6},
  pages={060501},
  year={2015},
  publisher={APS}
}

@article{saw2023configurationalA,
  title={Configurational temperature in active matter. I. Lines of invariant physics in the phase diagram of the Ornstein-Uhlenbeck model},
  author={Saw, Shibu and Costigliola, Lorenzo and Dyre, Jeppe C},
  journal={Physical Review E},
  volume={107},
  number={2},
  pages={024609},
  year={2023},
  publisher={APS}
}

@article{saw2023configurationalB,
  title={Configurational temperature in active matter. II. Quantifying the deviation from thermal equilibrium},
  author={Saw, Shibu and Costigliola, Lorenzo and Dyre, Jeppe C},
  journal={Physical Review E},
  volume={107},
  number={2},
  pages={024610},
  year={2023},
  publisher={APS}
}

@article{powles2005temperatures,
  title={Temperatures: old, new and middle aged},
  author={Powles, JG and Rickayzen, G and Heyes*, DM},
  journal={Molecular Physics},
  volume={103},
  number={10},
  pages={1361--1373},
  year={2005},
  publisher={Taylor \& Francis}
}

@article{hecht2024define,
  title={How to define temperature in active systems?},
  author={Hecht, Lukas and Caprini, Lorenzo and L{\"o}wen, Hartmut and Liebchen, Benno},
  journal={The Journal of Chemical Physics},
  volume={161},
  number={22},
  year={2024},
  publisher={AIP Publishing}
}

@article{berthier2017active,
  title={How active forces influence nonequilibrium glass transitions},
  author={Berthier, Ludovic and Flenner, Elijah and Szamel, Grzegorz},
  journal={New Journal of Physics},
  volume={19},
  number={12},
  pages={125006},
  year={2017},
  publisher={IOP Publishing}
}

@article{cugliandolo2019effective,
  title={Effective temperature in active Brownian particles},
  author={Cugliandolo, Leticia F and Gonnella, Giuseppe and Petrelli, Isabella},
  journal={Fluctuation and Noise letters},
  volume={18},
  number={02},
  pages={1940008},
  year={2019},
  publisher={World Scientific}
}

@article{burkholder2019fluctuation,
  title={Fluctuation-dissipation in active matter},
  author={Burkholder, Eric W and Brady, John F},
  journal={The Journal of chemical physics},
  volume={150},
  number={18},
  year={2019},
  publisher={AIP Publishing}
}

@article{caprini2021fluctuation,
  title={Fluctuation--dissipation relations in active matter systems},
  author={Caprini, Lorenzo and Puglisi, Andrea and Sarracino, Alessandro},
  journal={Symmetry},
  volume={13},
  number={1},
  pages={81},
  year={2021},
  publisher={MDPI}
}

@article{boriskovsky2024fluctuation,
  title={The fluctuation--dissipation relation holds for a macroscopic tracer in an active bath},
  author={Boriskovsky, Dima and Lindner, Benjamin and Roichman, Yael},
  journal={Soft Matter},
  volume={20},
  number={40},
  pages={8017--8022},
  year={2024},
  publisher={Royal Society of Chemistry}
}

@article{winkler2015virial,
  title={Virial pressure in systems of spherical active Brownian particles},
  author={Winkler, Roland G and Wysocki, Adam and Gompper, Gerhard},
  journal={Soft matter},
  volume={11},
  number={33},
  pages={6680--6691},
  year={2015},
  publisher={Royal Society of Chemistry}
}

@article{fuchs1992primary,
  title={Primary relaxation in a hard-sphere system},
  author={Fuchs, Matthias and Hofacker, I and Latz, A},
  journal={Physical Review A},
  volume={45},
  number={2},
  pages={898},
  year={1992},
  publisher={APS}
}

@article{fuchs1999aspects,
  title={Aspects of the dynamics of colloidal suspensions: Further results of the mode-coupling theory of structural relaxation},
  author={Fuchs, Matthias and Mayr, Matthias R},
  journal={Physical Review E},
  volume={60},
  number={5},
  pages={5742},
  year={1999},
  publisher={APS}
}

@article{marin2020tetrahedrality,
  title={Tetrahedrality dictates dynamics in hard sphere mixtures},
  author={Mar{\'\i}n-Aguilar, Susana and Wensink, Henricus H and Foffi, Giuseppe and Smallenburg, Frank},
  journal={Physical Review Letters},
  volume={124},
  number={20},
  pages={208005},
  year={2020},
  publisher={APS}
}

@article{marin2019slowing,
  title={Slowing down supercooled liquids by manipulating their local structure},
  author={Mar{\'\i}n-Aguilar, Susana and Wensink, Henricus H and Foffi, Giuseppe and Smallenburg, Frank},
  journal={Soft Matter},
  volume={15},
  number={48},
  pages={9886--9893},
  year={2019},
  publisher={Royal Society of Chemistry}
}

@article{boattini2020autonomously,
  title={Autonomously revealing hidden local structures in supercooled liquids},
  author={Boattini, Emanuele and Mar{\'\i}n-Aguilar, Susana and Mitra, Saheli and Foffi, Giuseppe and Smallenburg, Frank and Filion, Laura},
  journal={Nature Communications},
  volume={11},
  number={1},
  pages={5479},
  year={2020},
  publisher={Nature Publishing Group UK London}
}

@article{arenas2025structure,
  title={Structure--dynamics decoupling in soft-colloid suspensions},
  author={Arenas-Gullo, Adri{\'a}n and Clara-Rahola, Joaqu{\'\i}m and Segr{\'e}, Phil N and Ruiz-Franco, Jos{\'e} and Fernandez-Nieves, Alberto},
  journal={Nature Communications},
  year={2025},
  publisher={Nature Publishing Group UK London}
}

@article{royall2015role,
  title={The role of local structure in dynamical arrest},
  author={Royall, C Patrick and Williams, Stephen R},
  journal={Physics Reports},
  volume={560},
  pages={1--75},
  year={2015},
  publisher={Elsevier}
}

@article{ruiz2021role,
  title={On the role of competing interactions in charged colloids with short-range attraction},
  author={Ruiz-Franco, Jos{\'e} and Zaccarelli, Emanuela},
  journal={Annual Review of Condensed Matter Physics},
  volume={12},
  number={1},
  pages={51--70},
  year={2021},
  publisher={Annual Reviews}
}

@article{toledano2009colloidal,
  title={Colloidal systems with competing interactions: from an arrested repulsive cluster phase to a gel},
  author={Toledano, Juan Carlos Fernandez and Sciortino, Francesco and Zaccarelli, Emanuela},
  journal={Soft Matter},
  volume={5},
  number={12},
  pages={2390--2398},
  year={2009},
  publisher={Royal Society of Chemistry}
}

@article{campbell2005dynamical,
  title={Dynamical arrest in attractive colloids: The effect of long-range repulsion},
  author={Campbell, Andrew I and Anderson, Valerie J and van Duijneveldt, Jeroen S and Bartlett, Paul},
  journal={Physical review letters},
  volume={94},
  number={20},
  pages={208301},
  year={2005},
  publisher={APS}
}

@article{cardinaux2011cluster,
  title={Cluster-driven dynamical arrest in concentrated lysozyme solutions},
  author={Cardinaux, Fr{\'e}d{\'e}ric and Zaccarelli, Emanuela and Stradner, Anna and Bucciarelli, Saskia and Farago, Bela and Egelhaaf, Stefan U and Sciortino, Francesco and Schurtenberger, Peter},
  journal={The Journal of Physical Chemistry B},
  volume={115},
  number={22},
  pages={7227--7237},
  year={2011},
  publisher={ACS Publications}
}

@article{klix2010structural,
  title={Structural and dynamical features of multiple metastable glassy states in a colloidal system with competing interactions},
  author={Klix, Christian L and Royall, C Patrick and Tanaka, Hajime},
  journal={Physical review letters},
  volume={104},
  number={16},
  pages={165702},
  year={2010},
  publisher={APS}
}

@article{bomont2024arrested,
  title={Arrested states in colloidal fluids with competing interactions: A static replica study},
  author={Bomont, Jean-Marc and Pastore, Giorgio and Costa, Dino and Muna{\`o}, Gianmarco and Malescio, Gianpietro and Prestipino, Santi},
  journal={The Journal of Chemical Physics},
  volume={160},
  number={21},
  year={2024},
  publisher={AIP Publishing}
}

@article{gulati2026dynamical,
  title={Bicontinuity in active phase separation},
  author={Gulati, Paarth and Zhao, Liang and Tateno, Michio and Saleh, Omar A and Dogic, Zvonimir and Marchetti, M Cristina},
  journal={arXiv preprint arXiv:2601.03221},
  year={2026}
}

@article{kuvcera2022actin,
  title={Actin--microtubule dynamic composite forms responsive active matter with memory},
  author={Ku{\v{c}}era, Ond{\v{r}}ej and Gaillard, J{\'e}r{\'e}mie and Gu{\'e}rin, Christophe and Th{\'e}ry, Manuel and Blanchoin, Laurent},
  journal={Proceedings of the National Academy of Sciences},
  volume={119},
  number={31},
  pages={e2209522119},
  year={2022},
  publisher={National Academy of Sciences}
}

\end{document}


\maketitle

\begin{figure}[htbp!]
    \centering    \includegraphics[width=0.6\textwidth]{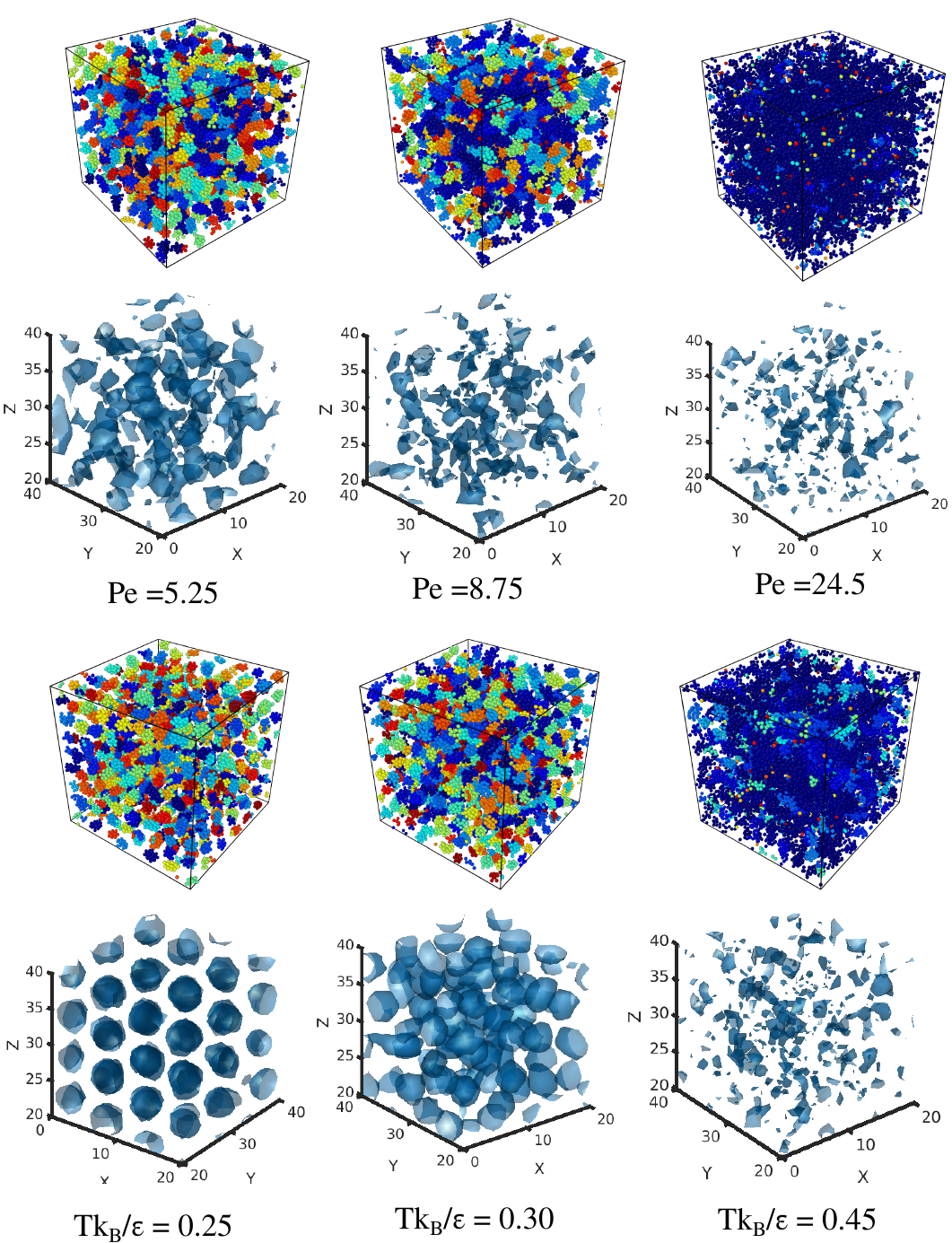}
    \caption{Typical steady state configurations and average positions PDF's of active and passive systems with equivalent structure for $\rho\sigma^3 = 0.155$ at three different $Pe-Tk_B/\epsilon$ pairs. The clusters of particles are colored differently for improving the visualization. The PDF's are represented with iso-probability surfaces. The iso-value for the first two columns is $2.5\times10^{-5}$ and $2.0\times10^{-5}$ for the last one. Although the snapshots look fairly the same for each pair, the PDF's already show important differences. In general, the passive system shows localized clusters at low temperatures, whereas the active system shows PDF's compatible with more mobile clusters. This is confirmed with the results discussed in the main text, where we demonstrate that at low $Pe$, although the internal structure of clusters is highly ordered, the active system exhibits diffusive behavior at long times.}
    \label{fig:ClustersS1}
\end{figure}

\begin{figure}[htbp!]
    \centering    \includegraphics[width=0.6\textwidth]{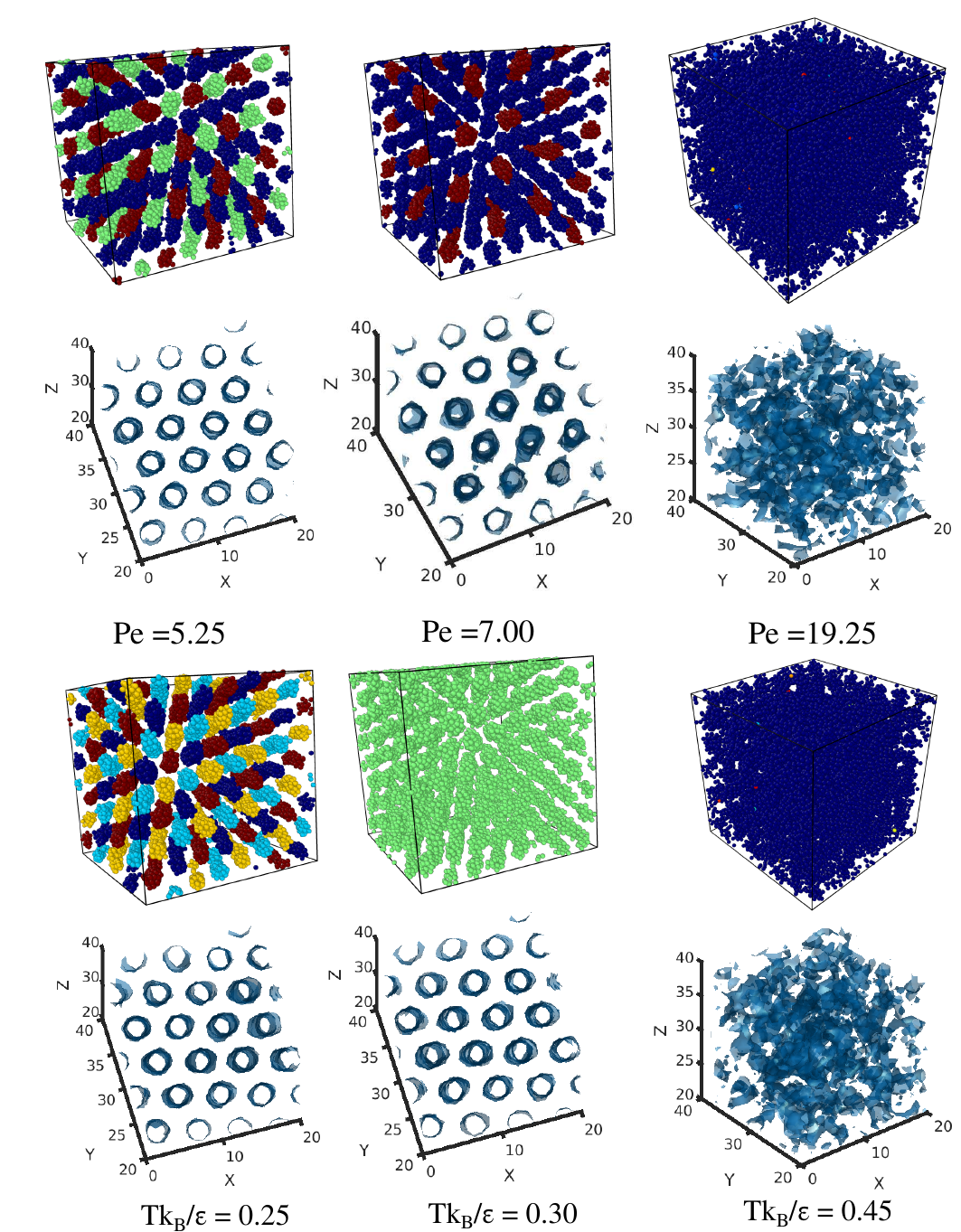}
    \caption{Typical steady state configurations and average positions PDF's of active and passive systems with equivalent structure for $\rho\sigma^3 = 0.264$ at three different $Pe-Tk_B/\epsilon$ pairs. The clusters of particles are colored in differently for improving the visualization The PDF's are represented with iso-probability surfaces. The iso-value for the first two columns is $2.5\times10^{-5}$ and $2.0\times10^{-5}$ for the last one. the phases look the same in both the active and passive system but exhibit different transport properties as discussed in the main text.}
    \label{fig:CylindersS2}
\end{figure}

\begin{figure}[htbp!]
    \centering    \includegraphics[width=0.6\textwidth]{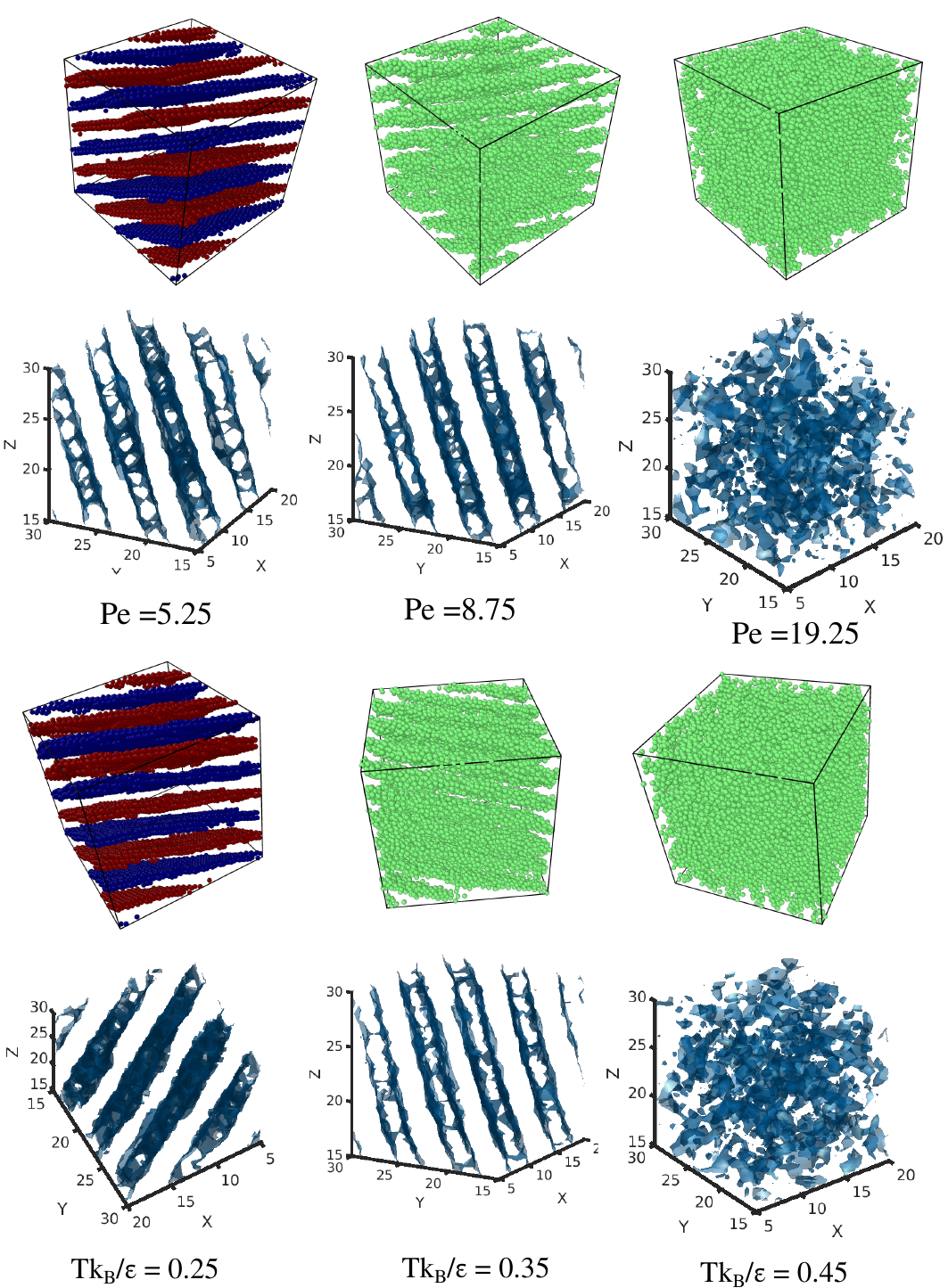}
    \caption{Typical steady state configurations and average positions PDF's of active and passive systems with equivalent structure for $\rho\sigma^3 = 0.418$ at three different $Pe-Tk_B/\epsilon$ pairs. The clusters of particles are colored in differently for improving the visualization The PDF's are represented with iso-probability surfaces. The iso-value for the first two columns is $2.5\times10^{-5}$ and $2.0\times10^{-5}$ for the last one. the phases look the same in both the active and passive system but exhibit different transport properties as discussed in the main text.}
    \label{fig:LamellarS3}
\end{figure}

\begin{figure}[htbp!]
    \centering    \includegraphics[width=0.95\textwidth]{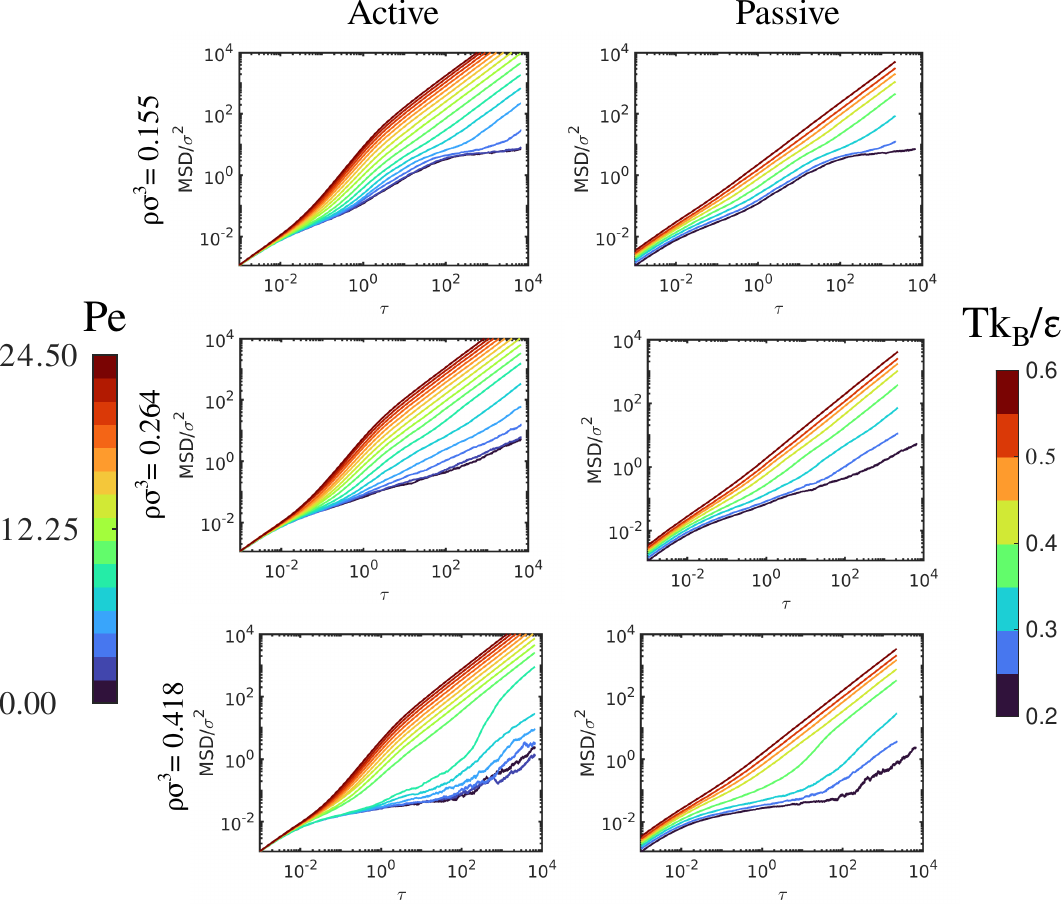}
    \caption{The mean squared displacement (MSD) of the systems considered in the article. The first column contain the results for the active system and the second for the passive. Each row corresponds to a given density as indicated.}
    \label{fig:MSD_S4}
\end{figure}

\begin{figure}[htbp!]
    \centering    \includegraphics[width=0.95\textwidth]{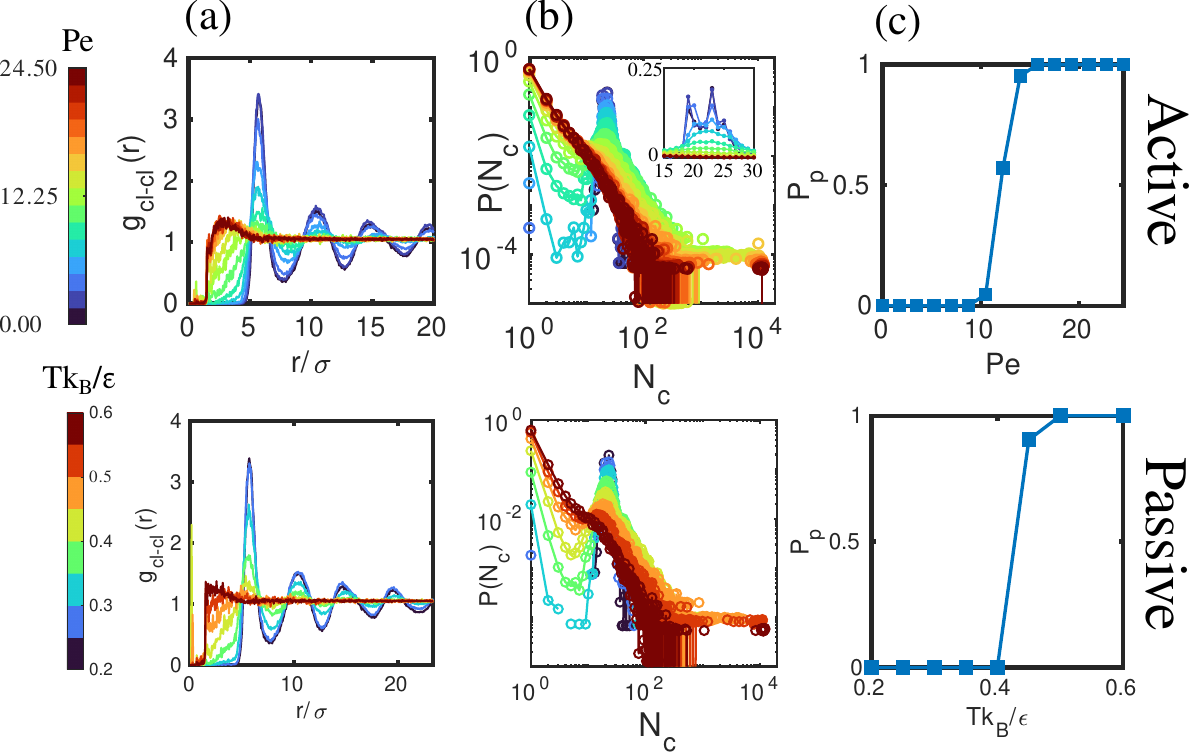}
    \caption{Comparison of structural properties of active and passive systems at $\rho\sigma^3 = 0.155$. The top row corresponds to the active system and the bottom row to the passive. \textbf{(a)} The cluster-cluster, pair correlation functions. \textbf{(b)} The cluster size distributions. The inset in the top panel shows three peaks at 19,23 and 26 particles per cluster for low $Pe$, corresponding to clusters with 2, 3 and 4 intertwined icosahedra as already discussed in \cite{serna2021formation}.  \textbf{(c)} The average steady probability of finding a percolating cluster. The similarity of the structural properties discussed in the main text and here is striking. Still, the transport properties are different when comparing the active and passive system, leading to a structure-dynamics decoupling as explained in the main text.}
    \label{fig:ClustersStructureS5}
\end{figure}

\begin{figure}[htbp!]
    \centering    \includegraphics[width=0.95\textwidth]{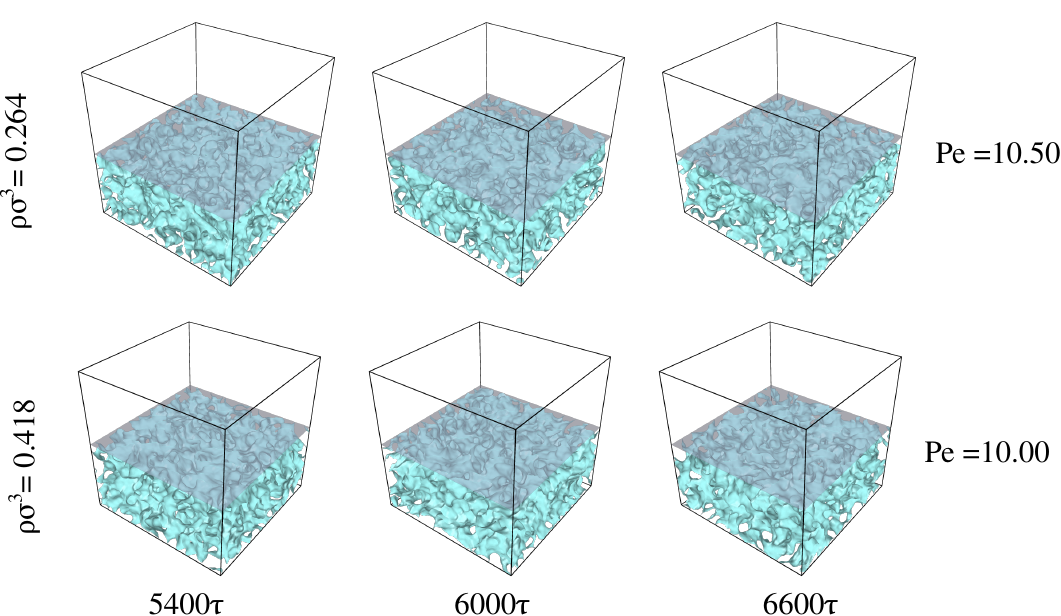}
    \caption{Instantaneous positions PDF of selected systems at the steady state exhibiting a disordered bicontinuous structure. The interconnected structures evolve in time but keep steady structural observables like $g(r)$ or local symmetry extracted from a CNA as described in the main text. Here we only present the structures in the active system, but the passive ones exhibit similar features at the temperatures above the stability regions of hexagonal cylindrical and lamellar phases. SALR systems are known to self-assemble ordered bicontinuous phases at equilibrium. Although studying the ordered bicontinuous phase (active or passive) is out of the scope of the present work, the recent interest on active bicontinuous phases \cite{gulati2026dynamical} makes this feature of SALR systems inspiring for future research.}
    \label{fig:Bicontinuous_S5}
\end{figure}

\begin{figure}[htbp!]
    \centering    \includegraphics[width=0.5\textwidth]{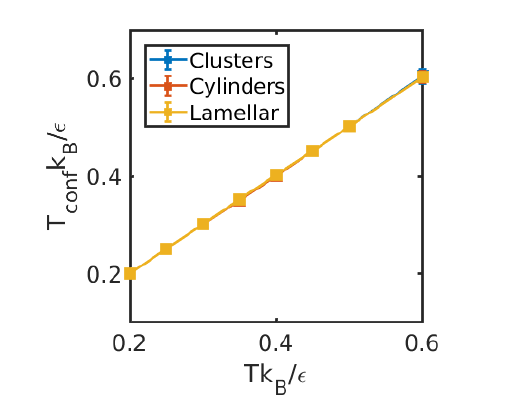}
    \caption{The configurational temperature, $T_{conf}k_B/\epsilon$, vs the temperature of the thermal bath, $Tk_B/\epsilon$, for passive systems at $\rho\sigma^3 = 0.155$ (Clusters), $\rho\sigma^3 = 0.264$ (Cylinders) and $\rho\sigma^3 = 0.418$ (Lamellar). As expected, both temperatures coincide perfectly.}
    \label{fig:T_conf_Passive_S7}
\end{figure}

\begin{figure}[htbp!]
    \centering    \includegraphics[width=0.35\textwidth]{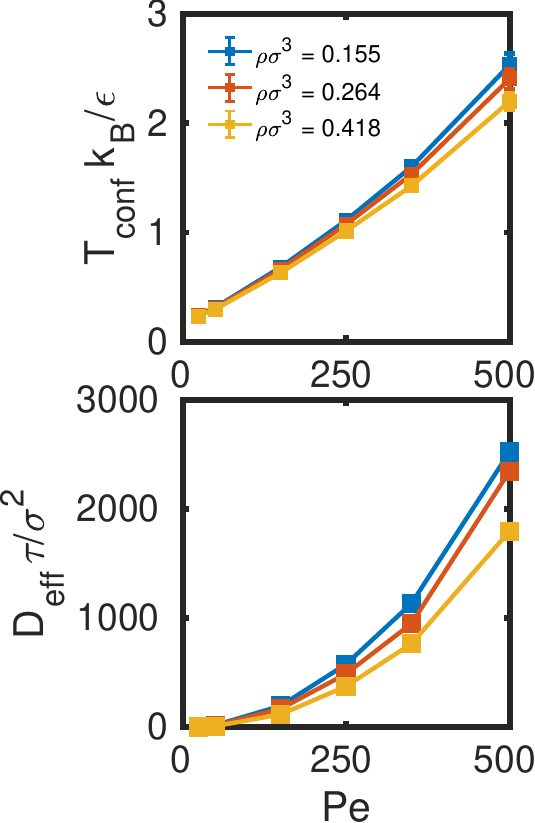}
    \caption{The configurational temperature, $T_{conf}k_B/\epsilon$, and the effective diffusion coefficient, $D_{eff}\tau/\sigma^2$ as a function of $Pe$ in the high activity regime. As discussed in the main text, the effective diffusion coefficient coincides with $T_{eins}$. Even at high activities $T_{conf}$ is almost independent of the density, with small variations very often within the error. However, $T_{eins}$ shows a strong density dependence especially at high $Pe$.}
    \label{fig:HighPe_S8}
\end{figure}

\begin{figure}[htbp!]
    \centering    \includegraphics[width=0.35\textwidth]{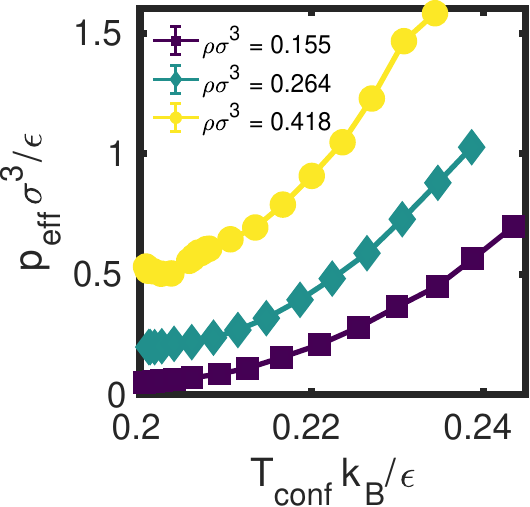}
    \caption{The non-equilibrium equation of state as $p_{_\text{eff}}\sigma^3/\epsilon$ vs $T_\text{conf}k_B/\epsilon$. As for the density dependence, it keeps the same trend as in the passive case when increasing the temperature of the thermal bath, following a general increase of $p_{_\text{eff} }\sigma^3/\epsilon$ as $\rho\sigma^3$ increases. The solid-like lamellar phase ($\rho\sigma^3 = 0.418$, at low $T_\text{conf}k_B/\epsilon$), presents a decrease of $p_{_\text{eff}}\sigma^3/\epsilon$ that correlates with a more energetically stable structure. The transitions are signaled in the equation of state as changes of slope and small jumps, being the most evident those of the densest system}
    \label{fig:EoS_S9}
\end{figure}

\bibliography{apssamp}